\documentclass[12pt]{article}
\usepackage{graphicx}
\usepackage{color}
\usepackage{multirow}
\usepackage{amsmath}
\usepackage{booktabs}
\usepackage{dcolumn}
\usepackage{rotating}

\definecolor{MyGreen}{cmyk}{1.0,0.0,1.0,0.2}

\definecolor{MyFuschia}{cmyk}{0.07,0.95,0,0}

\usepackage{scicite}

\newcounter{lastnote}

\usepackage{times}

\newenvironment{sciabstract}{%
\begin{quote} \bf}
{\end{quote}}

\title{Kantian Fractionalization Predicts the Conflict Propensity of the International System}

\date{}

\begin{document}

\baselineskip 24pt

\maketitle

\vspace*{-1in}
\begin{center}
Skyler J. Cranmer$^{1,\ast}$, Elizabeth J. Menninga,$^1$ and Peter J. Mucha$^2$\\ \bigskip
\normalsize{$^1$Department of Political Science, University of North Carolina, Chapel Hill, NC, USA}\\
\normalsize{$^2$Department of Mathematics, University of North Carolina, Chapel Hill, NC, USA}\\ \bigskip
\normalsize{$^\ast$To whom correspondence should be addressed; E-mail: skyler@unc.edu.}\\
\end{center}

\begin{sciabstract}
The study of complex social and political phenomena with the perspective and methods of network science has proven fruitful in a variety of areas \cite{Kossinets:2006,Onnela:2007,Lazer:2009,Centola:2010,Bassett:2011}, including applications in political science\cite{Porter:2005,Fowler:2010,MuchaPorter:2010} and more narrowly the field of international relations \cite{Maoz:1993,Maoz:2006,Ward:2007}.
We propose a new line of research in the study of international conflict by showing that the multiplex fractionalization of the international system (which we label \emph{Kantian fractionalization}) is a powerful predictor of the propensity for violent interstate conflict, a key indicator of the system's stability. In so doing, we also demonstrate the first use of multislice modularity\cite{Mucha:2010} for community detection in a multiplex network application.
Even after controlling for established system-level conflict indicators, we find that Kantian fractionalization contributes more to model fit for  violent inter-state conflict than previously established measures. Moreover, evaluating the influence of each of the constituent networks shows that joint democracy plays little, if any, role in predicting system stability, thus challenging a major empirical finding of the international relations literature. Lastly, a series of Granger causal tests shows that the temporal variability of Kantian fractionalization is consistent with a causal relationship with the prevalence of conflict in the international system. This causal relationship has real-world policy implications as changes in Kantian fractionalization could serve as an early warning sign of international instability. 
\end{sciabstract}

\noindent \textbf{One Sentence Summary:} Network fractionalization powerfully predicts the stability of the international system, casting doubt on the most prominent finding in the study of conflict. \\

Immanuel Kant proposed a recipe for international peace in 1795\cite{Kant:1795} that has proven remarkably insightful: diffusion of democracy, economic interdependence, and establishment of international institutions.  Past studies in international relations have explored the impacts of the components of the Kantian tripod individually\cite{Barbieri:1996, Maoz:1993, Russett:1990} as well as collectively \cite{Bennett:2003, Russett:2001} on the prospects of peace. These studies, however, include democracy, interdependence, and intergovernmental organizations (IGOs) as three  independent variables in regressions in which the outcome is a measure of dyadic war.  This approach is limited insofar as each component is inherently relational and thus has implications for the entire international system, not just pairs of states. Joint democracy describes a similarity between two states that connects them politically. IGOs link member countries together through common norms, principles, and procedures. Trade interconnects the economic growth and stability of states.   Some scholars have considered the effects of system level measures of the Kantian tripod\cite{Maoz:2011, OnealRussett:1999} but the outcome of interest is still dyadic. A few studies do consider conflict at the system level\cite{GleditschHegre:1997, CrescenziEnterline:1999}, but these studies  look at the effect of democracy alone, ignoring the other components of the Kantian tripod. 

We believe that to  evaluate the effect of Kant's prescription for peace on international conflict, the three components of the Kantian tripod must be considered collectively. Moreover, as dyadic relations are influenced by, and influence, the other relationships in the system, Kant's prescription for peace should not merely be applied to  dyadic conflict; it should have implications for conflict at the system level.  In short, we improve upon past international relations studies in two ways:  quantifying the nature of interconnectedness of the international system by combining the pieces of the Kantian tripod together at the system level in our \emph{Kantian fractionalization} measure and considering the effect of \emph{Kantian fractionalization} on systemic measures of conflict.



The fundamental idea behind the Kantian peace is that the more interconnected the international system becomes, the less likely conflict is to occur. As democracy spreads, states become more economically interdependent, and IGOs grow in scope and power,  war becomes more costly and alternatives to war become both more abundant and more appealing.   Democracy checks executive power and promotes norms of compromise and negotiation.  Trade increases the stakes of the conflict and provides incentives to resolve disputes without damaging mutually beneficial relationships.  IGOs present forums and procedures for peaceful conflict resolution. Likewise, when these factors become less prevalent and the connectivity among states weakens, credible alternatives to war become harder to find and thus the potential for violent conflict increases. 

The level and organization of interconnectivity in the international system is indicative of the level of stress being exerted on relationships between states. Connections that encompass more states with lower levels of fractionalization induce less stress on the international system. When a dispute arises between states, both the dyadic and the system-level connections are relevant. Broad-based system-level political and economic connections through trade, IGOs, or democracy provide avenues for non-violent conflict resolution strategies. Additionally, these connections increase the incentives for states to identify peaceful solutions to their disputes, and thus the likelihood of war is decreased.  States will still find themselves in disagreements with one another, but these disputes are less likely to escalate to war in a system with low fractionalization.  When fractionalization in the international system is high, however, greater tension is exerted on the international system. Conflicts that arise in very tense systems are more likely to escalate to the use of military force as the collective system-level pacifying effects of democracy, trade, and IGOs are too weak to mitigate this tension.  We refer to the level of division in the international system as the system's \emph{Kantian fractionalization}, and we expect higher levels of Kantian fractionalization will result in higher incidence of interstate conflicts.



In order to measure the level of fractionalization of the international system, we utilize the tools of community detection in networks \cite{Porter:2009,Fortunato:2010}. A community in a network is a group of vertices that are more strongly connected to one another than they are to the rest of the network, building on classic ideas of graph partitioning from computer science and cohesive groups from social science. One of the dominant methods of community detection has been the computational optimization of  modularity \cite{Newman:2004}. 
Modularity is a direct quantification of the notion that a good partition of the network's vertices into communities (wherein every vertex receives an assignment to one and only one of the identified communities) identifies groups of vertices that are more tightly connected to each other than to the rest of the network. Specifically, modularity is calculated as the total weight of intra-group edges minus the expected number under an appropriate null model.
 Larger modularity values, therefore, signal denser, stronger connections between vertices in the same community relative to the network as a whole, with relatively sparser, weaker connections between communities. 

We quantify the concept of the Kantian tripod as a network between states (vertices) with edges describing the weights of ties between states in (directed) trade, joint IGO memberships, and joint democracy status. We then measure the fractionalization by \emph{multislice modularity} \cite{Mucha:2010}.  Because of the multiple kinds of between-state ties in the Kantian tripod (a \emph{multiplex} network), we use multislice modularity in its multiplex network form, treating each of the three kinds of connections---trade, common IGO membership, and joint democracy---as a slice of the multiplex network, considering each year of data separately. In this formulation of the multiplex Kantian network, each state is represented as three (multislice) vertices that are connected to one another by identity arcs of weight specified by the interslice coupling parameter, $\omega$. The community detection is then performed by a computational heuristic\cite{GenLouvain} that partitions the (multislice) vertices  into communities to  maximize the obtainable value of multislice modularity, $Q_K$. 

Previous applications of multislice modularity have successfully contributed to the study of longitudinal network data, including correlations in legislative voting patterns \cite{MuchaPorter:2010}, brain activity \cite{Bassett:2011}, and behavioral similarity over time \cite{Wymbs:2012}. To our knowledge, this is the first use in practical application of multislice modularity in a multiplex network [that is, separate from the limited in-principle demonstration that accompanied the original development of multislice modularity\cite{Mucha:2010}]. As such, we take particular care to scale the weights of the three slices (or `layers') accordingly and investigate the effects of our parameter choices to test the robustness of our results (see Supporting Online Material [SOM] for details). While in principle the same multislice modularity methodology can be applied to data that is both multiplex and longitudinal, such consideration would introduce additional parameters and complexity beyond the scope of the current contribution; moreover, as we show below, community detection on single-year multiplex data provides us with a suitable measure of fractionalization at each time point. 

As a direct extension of modularity, multislice modularity inherits the established positive attributes of modularity, including intuitions developed through its broad use and a wide array of available computational heuristics for its optimization. Of course, multislice modularity also shares the well-known drawbacks of the original modularity formulation, including its resolution limit\cite{Fortunato:2007} and the presence of many near-optimal partitions \cite{Good:2010}. The sizes of identified communities is influenced by a spatial resolution parameter $\gamma$ \cite{Reichardt:2006}.  We consider multiple values of the spatial resolution parameter and identity coupling parameters to ensure our results are not sensitive to these specifications. The possible existence of many nearly-optimal partitions of the network into communities does not affect our results in the present work, as we do not consider the particular assignments of vertices into communities in our analysis (although we do include visualizations of the community assignments in the SOM). Indeed, in utilizing only the value of Kantian fractionalization, the possible presence of many nearly-optimal partitions of the network gives greater confidence in the computationally obtained $Q_K$ values. For extra confidence, we consider many realized outputs from the selected heuristic\cite{GenLouvain} and considered various parameter specifications (as described in the SOM).

To calculate Kantian fractionalization, we first quantified the connections in the multiplex international network in each year with respect to the three principal aspects of the Kantian peace \cite{Russett:2001}: joint democracy, trade, and common IGO membership. The joint democracy layer is defined to be a clique (of unit edge weight) connecting all democracies (states with a Polity IV\cite{Marshall:2002} score greater than 6, a standard threshold for democracy in international relations), leaving non-democracies isolated (except for their interslice identity connections to the other layers). The trade layer is a directed network with non-zero edge weights linearly related to the logarithm of trade value from one country to another. The IGO network layer is defined as an undirected weighted network with the edge weight connecting two states proportional to the number of common IGO memberships. We scaled the trade and IGO layers to have median present edge-weight equal to 1 so that the weight distributions of the three layers would be qualitatively similar, as detailed in the SOM.  The Kantian fractionalization  of the international system in a given year is then defined as the maximum obtained value of
\[
Q_K = \frac{1}{2\mu}\sum_{ijlr}\left\{\left(A_{ijl}-\gamma P_{ijl} \right)\delta_{lr} + \delta_{ij}(1-\delta_{lr})\omega\right\} \delta(g_{il},g_{jr})\,,
 \]
where $A_{ijl}$ is the edge weight connecting states $i$ and $j$ in layer $l$, $P_{ijl}$ is the corresponding null model in layer $l$ (Newman-Girvan\cite{Newman:2004} for IGO and joint democracy, Leicht-Newman\cite{Leicht:2008} for trade), $\gamma$ is a spatial resolution parameter, $\omega$ is the specified interslice identity coupling, $g_{il}$ is the community assignment of vertex $i$ in layer $l$, and Kronecker $\delta$ indicators equal $1$ when their two arguments are identical ($0$ otherwise). For our principal $Q_K$ specification, we use the default values $\gamma=\omega=1$. In order to have confidence in the obtained $Q_K$ values, we run the selected computational heuristic \cite{GenLouvain} 100 times with pseudorandom vertex orders and select the maximum observed value. To further ensure the robustness of our results to the selected parameter values, we explore alternative choices in the SOM. 


Having established our measure of system interconnectedness, we explore the relationship between these yearly Kantian fractionalization values and the quantity of violent conflict in the international system. We operationalize international conflict by examining the number of times in a calendar year violent military force is ``explicitly directed towards the government, official representatives, official forces, property, or territory of another state'' \cite[p.163]{Jones:1996}. This type of conflict is generally called a militarized interstate dispute, and we include disputes marked by violence ranging in intensity from small skirmishes to full scale war. Often, such disputes last several years, but because we are interested in system stability, we restrict our focus to the onset of new violent conflicts. Simply counting the onset of conflicts, however, fails to account for the fact that during our period of observation, 1949--2000, the number of states in the system increases from 72 to 191, providing more opportunities for dyadic interstate conflict \cite{Cranmer:2012}. As such, our outcome variable, \emph{conflict rate}, is measured as the density of new violent conflicts in the available dyads in the international system-year. To assure robustness, we also consider a second measure of \emph{conflict rate} that is the direct count of new violent conflicts in our count models with an explicit adjustment for the number of dyads.

In our statistical analyses, we lag the modularity measures  one year to account for the fact that a causal relationship between Kantian fractionalization and conflict implies the temporal precedence of Kantian fractionalization. We also control for several other variables that are common in the international conflict literature to capture factors related to system stability \cite{Maoz:2005,Maoz:2006}. First, we include Moul's measure of system polarity.\cite{Moul:1993} This measure divides the number of major power alliance groups by the number of major powers, thus producing a ratio to capture the polarity of the international system. We also include lagged (one year) defensive alliance interdependence as measured by Maoz \cite{Maoz:2006}. To account for the role that the distribution of material capabilities are traditionally thought to play in system stability, we include a five-year rolling average of movement in capability concentration using Ray and Singer's measure \cite{Ray:1973}. Finally, we include a one-year lagged outcome variable to account for the first-order autocorrelation observed in the outcome variables. See the SOM for details on the measurement of the controls and for the establishment of first-order autocorrelation.

\begin{figure}[t]
\centering
\includegraphics[width=8cm]{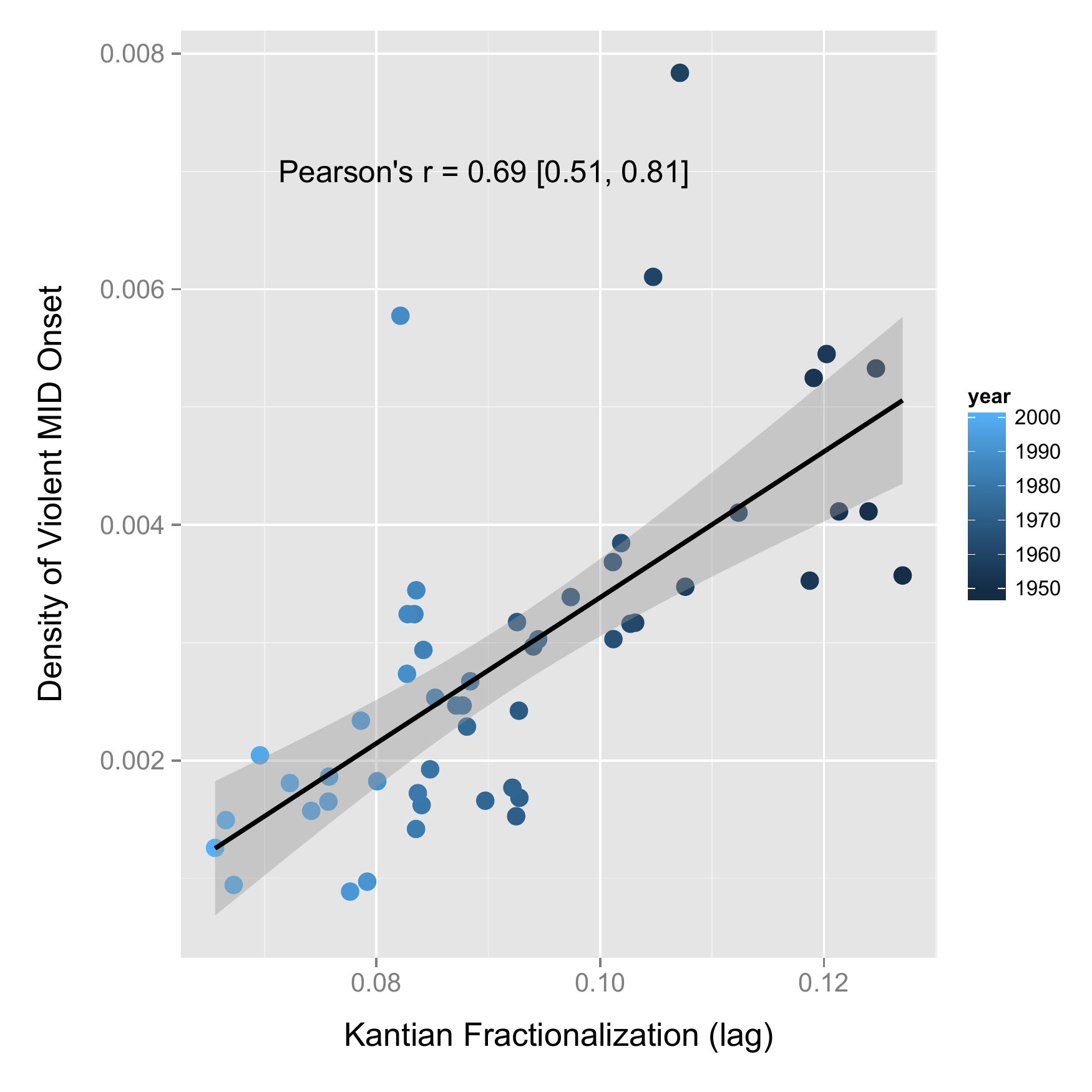}
\caption{Kantian fractionalization, lagged by one year, versus conflict rate, 1949-2000. The line and confidence bands reflect those fit by a simple bivariate linear model.}
\label{scatter}
\end{figure}

The observed bivariate relationship between Kantian fractionalization and conflict rate is strong. Figure \ref{scatter} shows Kantian fractionaliation, lagged by one year, plotted against conflict rate, a clear and apparently linear relationship exists between the two. As Kantian fractionalization increases, so does the rate of conflict. The graphical relationship is  born out statistically ($r$ = 0.690, $p < 0.001$). Given the clear bivariate relationship, and the trends of both variables with time, we must satisfy ourselves that the relationship holds in the presence of the aforementioned controls. Table \ref{reg_table} shows the results of linear regressions of conflict rate and Poisson regressions of the direct count of new conflicts per year, offset by the opportunities for conflict (the log of number of dyads in the system year) so that the model captures the rate of conflict \cite{Gelman:2007}. Furthermore, the count models include a dispersion parameter to adjust the standard errors for the over-dispersion present in the annual count of violent conflicts \cite{Gelman:2007}. For both the linear and Poisson regressions, we compare a simple specification with only our fractionalization measure and a lagged outcome variable to a model that includes all of the controls discussed above. 

\begin{table}
\begin{scriptsize}
\noindent\makebox[\textwidth]{%
\begin{tabular}{rrrrrrrrrrrrr}
& \multicolumn{6}{c}{Linear Models of Conflict Onset Density} & \multicolumn{6}{c}{Count Models of Conflict Onset}\\
  & \multicolumn{2}{c}{Basic} & \multicolumn{2}{c}{With Controls} & \multicolumn{2}{c}{Without Frac.} & \multicolumn{2}{c}{Basic} & \multicolumn{2}{c}{With Controls} & \multicolumn{2}{c}{Without Frac.} \\ 
  \hline
Kantian Fract. (lag) & \textbf{0.041} & \textbf{(0.012)} & \textbf{0.055} & \textbf{(0.019)} &  &  & \textbf{24.143} & \textbf{(3.394)} & \textbf{24.817} & \textbf{(5.696)} &  &  \\ 
  Moul Polarity &  &  & -0.000 & (0.001) & -0.001 & (0.001) &  &  & -0.167 & (0.202) & \textbf{-0.726} & \textbf{(0.165)} \\ 
  Alliance Dep. (lag) &  &  & 0.007 & (0.004) & \textbf{0.003} & \textbf{(0.004)} &  &  & 2.169 & (1.220) & 1.624 & (1.455) \\ 
  Sys. Movement &  &  & 0.004 & (0.018) & 0.029 & (0.017) &  &  & 0.332 & (6.690) & \textbf{15.177} & \textbf{(6.147)} \\ 
  Lagged Outcome & \textbf{0.347} & \textbf{(0.131)} & 0.220 & (0.146) & \textbf{0.464} & \textbf{(0.129)} & \textbf{0.018} & \textbf{(0.006)} & \textbf{0.013} & \textbf{(0.006)} & 0.006 & (0.007) \\ 
  (Intercept) & \textbf{-0.002} & \textbf{(0.001)} & -0.007 & (0.003) & 0.001 & (0.002) & \textbf{-8.520} & \textbf{(0.366)} & \textbf{-9.553} & \textbf{(1.072)} & \textbf{-5.891} & \textbf{(0.804)} \\
   \hline
   Adjusted $R^2$ / AIC & 0.523 & & 0.531 & & 0.455 & & 366.54 & & 364.18 & & 406.67 & \\
\end{tabular}}
\caption{Linear models of violent conflict onset density and count models of violent conflict onset in the interstate system. Count models correct for over-dispersion and include the log of the number of dyads in the system year as an offset. Coefficients and standard errors displayed in bold are statistically significant at or below the $p=0.05$ level.}
\label{reg_table}
\end{scriptsize}
\end{table}

The results in Table \ref{reg_table} show that Kantian fractionalization consistently maintains a statistically significant and substantively large positive effect on the occurrence of conflict, regardless of specification. The models with Kantian fractionalization also display consistently superior in-sample fit. Indeed, the simple specifications containing only Kantian fractionalization and a lagged outcome variable have higher $R^2$ and lower AIC statistics than models with all controls, but without Kantian fractionalization. These results are also substantively robust to the measurement of modularity. We conducted similar analyses with Kantian fractionalization operationalized by other choices of the $\gamma$ and $\omega$ parameters, yielding qualitatively identical conclusions (see SOM for details). Furthermore, we computed modularity using only trade and only IGO connections respectively (see the SOM) and found that those models did not fit as well as the full Kantian fractionalization models. Lastly, likelihood ratio tests reveal that restricting both the linear and count models with controls to exclude Kantian fractionalization is an invalid restriction ($\chi^2 = 8.928$, $p=0.003$ for the linear model; $\chi^2=44.495$, $p=2.55\cdot 10^{-11}$ for the count model).

\begin{figure}[t]
\centering
\begin{tabular}{cc}\vspace{-0.5cm}
\includegraphics[width=7cm]{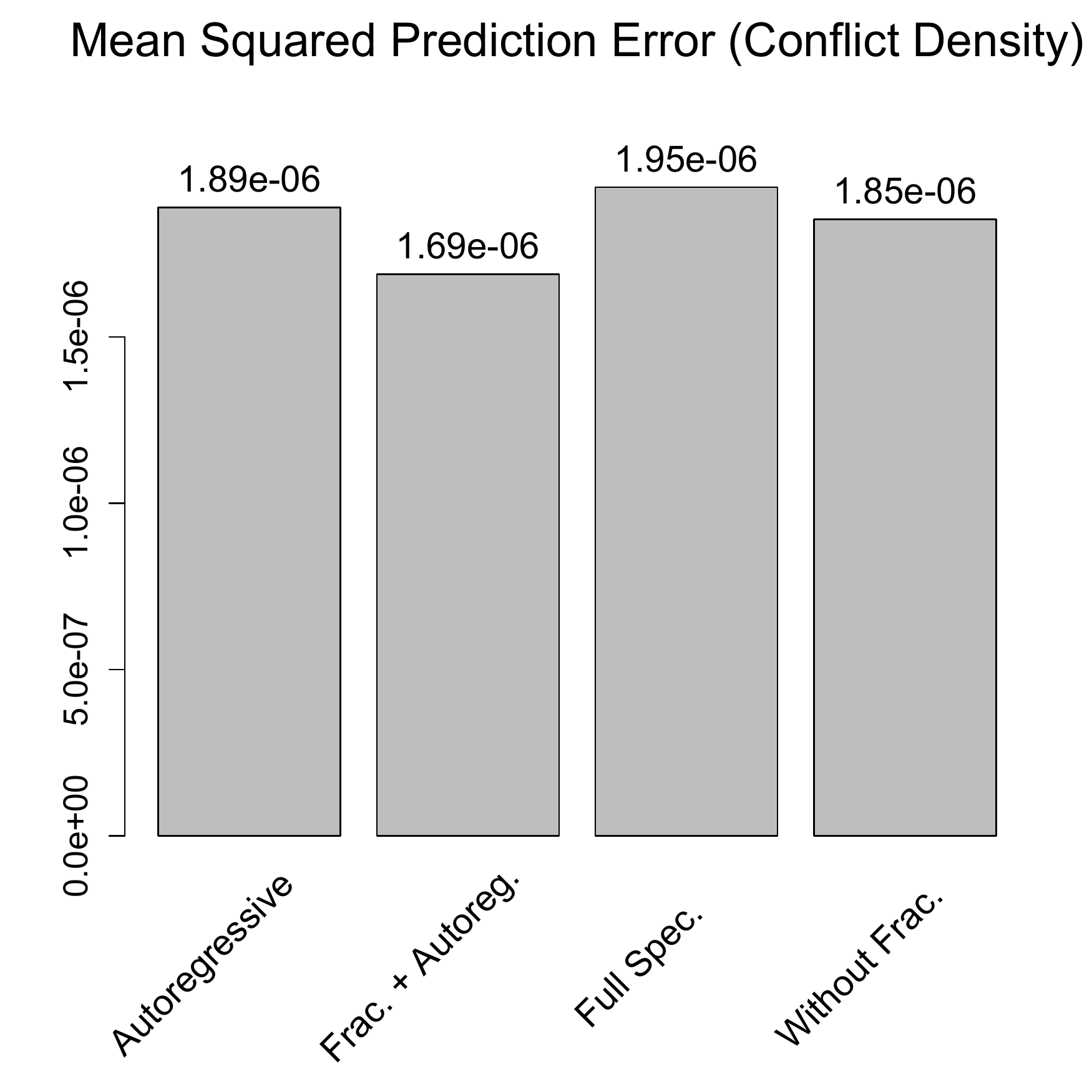}&
\includegraphics[width=7cm]{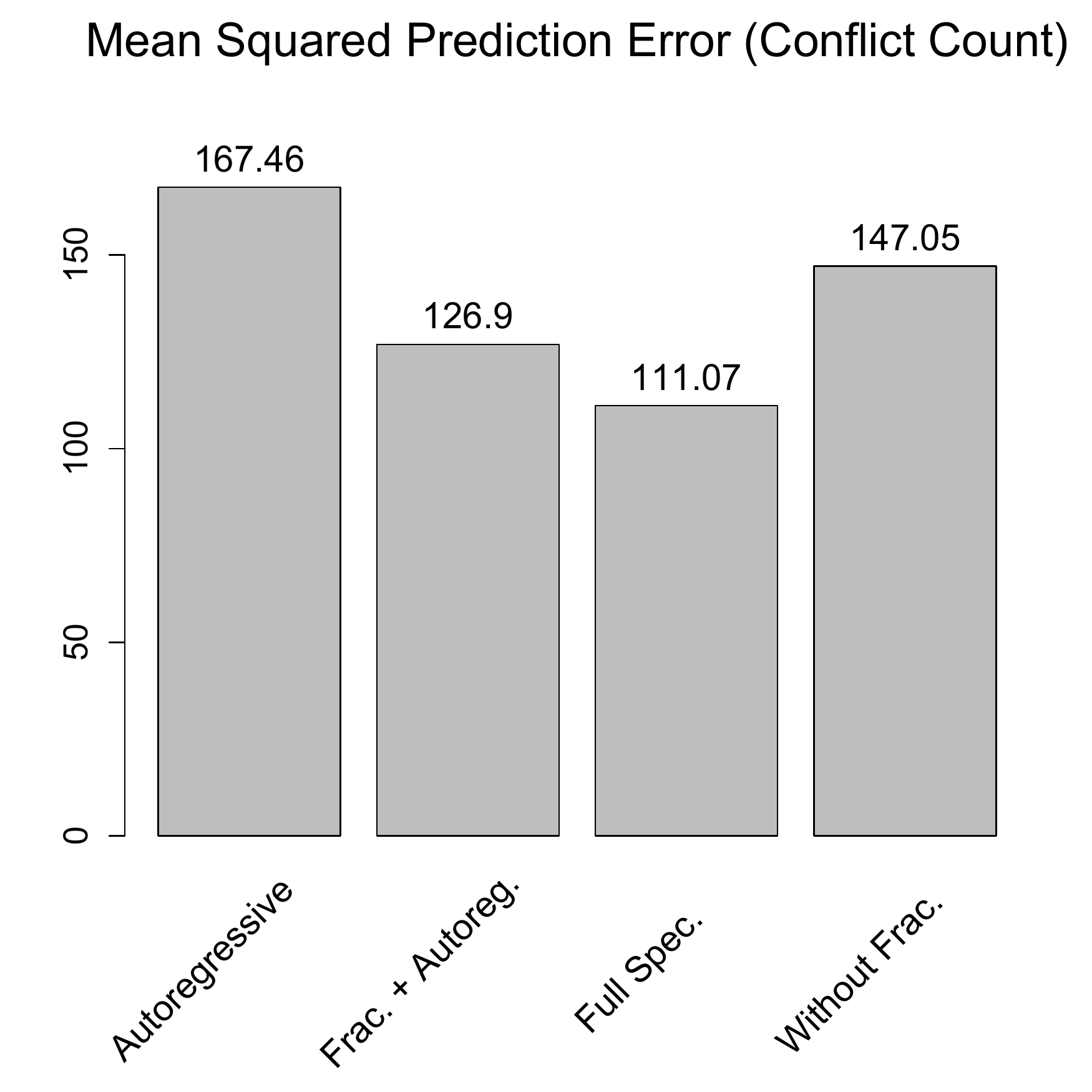}
\end{tabular}
\caption{Out-of-sample (one year ahead) predictive performance. Both plots show the mean squared prediction error from a series of forecasts in which the values of both conflict density and conflict count, the left and right plots respectively, were forecast using only the data available up to, but not including, the year forecast.}
\label{prediction}
\end{figure}

We further consider the predictive power of our measure and model. We forecast the level of conflict expected in year $t$ by estimating a model on the data ranging from the beginning of our time window until time $t-1$ and fitting those coefficients to the data at time $t$.  Figure \ref{prediction} shows that adding our Kantian fractionalization measure to the model always improves the out-of-sample predictive fit.


The correlation tests and regressions do not, however, address the issue of whether Kantian fractionalization causes conflict, as our theory  suggests, or conflict causes Kantian fractionalization. Indeed, it is theoretically feasible that conflict could cause the severing of trade and IGO connections that would result in higher Kantian fractionalization. To address the possibility of reverse effects, and satisfy ourselves to the greatest extent possible that the relationship between Kantian fractionalization and conflict is consistent with a causal interpretation, we conducted a series of Granger-causal tests. Some variable $x$ is said to Granger cause another variable $y$ if lagged values of $x$ are statistically reliable predictors of current values of $y$, but the reverse is not true \cite{Granger:1969}. Note that Granger causality does not capture causality in the potential outcomes sense \cite{Rubin:1974}, but shows consistency with the causal story in terms of temporal dynamics. 

\begin{table}[t]
\begin{scriptsize}
\begin{center}
\begin{tabular}{c|cc|cc}
  \hline
& \multicolumn{2}{c}{Conflict Dens. $\rightarrow$ Kantian Frac.}\vline
& \multicolumn{2}{c}{Kantian Frac. $\rightarrow$ Conflict Dens.}\\
Lags & $F$-Statistic & $p$-Value & $F$-Statistic & $p$-Value\\
\hline
1 & 0.024 & 0.877 & \textbf{12.123} & \textbf{0.001} \\ 
  2 & 0.101 & 0.904 & \textbf{4.449} & \textbf{0.017} \\ 
  3 & 0.229 & 0.876 & \textbf{3.863} & \textbf{0.016} \\ 
  4 & 0.901 & 0.473 & \textbf{4.375} & \textbf{0.005} \\ 
  5 & 0.677 & 0.644 & \textbf{3.547} & \textbf{0.010} \\ 
  6 & 0.428 & 0.855 & \textbf{2.409} & \textbf{0.048} \\ 
  7 & 0.700 & 0.672 & \textbf{3.044} & \textbf{0.015} \\ 
  8 & 0.898 & 0.531 & 2.036 & 0.079 \\ 
  9 & 0.610 & 0.777 & 1.260 & 0.306 \\ 
  10 & 0.608 & 0.790 & 1.046 & 0.440 \\    \hline
\end{tabular}
\caption{Granger causal analysis of violent conflict onset density and Kantian fractionalization. The $F$-statistics and $p$-values shown in bold are statistically significant at or below the $p=0.05$ level.}
\label{granger_table}
\end{center}
\end{scriptsize}
\end{table}

Table \ref{granger_table} shows the Granger causal tests between one and ten year lags. When Kantian fractionalization  is lagged between one and seven years, it predicts current values of conflict rate, but lagged values of conflict rate do not predict current values of Kantian fractionalization. As such, we conclude that Kantian fractionalization  Granger causes conflict rate with up to seven year lags. The statistical significance of this effect drops off for lags longer than seven years.

\begin{figure}[t]
\centering
\includegraphics[width=8cm]{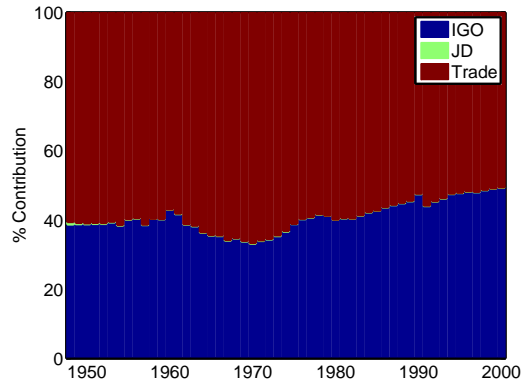}
\caption{Relative contribution of the three networks to our Kantian fractionalization measure. We note in particular that the three slices of the Kantian tripod employed here (IGO, joint democracy, trade) have been scaled so that the median present edge weights are the same in each (see the SOM for details).  Nevertheless, the numbers, relative weights, and patterns of connections are such that the relative contributions to Kantian fractionalization are dominated by IGO and trade, with little contribution from joint democracy.}
\label{Qcontrib}
\end{figure}

We can also examine the relative contribution of the three network layers---trade, joint IGO membership, and joint democracy---to our Kantian fractionalization measure in order to judge which of the layers are most central to the measure. When setting the initial relative weights for our computation of the Kantian fractionalization, we decided on a principle of equal effects of the three layers on the dyads to scale the present edge weights in each layer so that the full temporal distribution of each type has unit median. However, equal median edge weights in each layer do not necessarily induce equal impact on multislice modularity, not only because of variation over time but also because the total number of edges varies and the clustering of edge weights may be qualitatively different from one layer to another. To quantify the contribution from each network layer, we compare Kantian fractionalization to multislice modularity values obtained by permuting country identities and calculate the mean increase in modularity from each layer.  Figure \ref{Qcontrib} shows that the  majority of the measure (as identified in this permutations-based manner) is driven by trade and IGO connections, whereas joint democracy plays little role at all. (Indeed, as the SOM shows, our results do not change in a meaningful way if we drop joint democracy from the computation entirely.) This is a noteworthy result as the  tendency for jointly democratic dyads not to engage in militarized conflict is arguably the most significant and heretofore robust empirical finding of the last several decades of scholarship on international politics. Our result suggests that the idea of a democratic peace, while generally thought credible at the dyadic level, does not scale up to become a meaningful predictor of system stability.

Taken together, our results suggest that (a) a relationship between Kantian fractionalization and conflict in the international system exists, (b) the correlation seems not to be spurious, (c) our Kantian fractionalization measure does more to improve the out-of-sample predictive fit of the model than any other measure existent in the literature, (d) the temporal dynamics of the relationship are consistent with a causal effect, and (e) the composition of our measure casts doubt on the system-level influence of a democratic peace in international relations. These results are robust across multiple operationalizations of Kantian fractionalization, multiple model specifications, and multiple statistical models.  

We have introduced both a new way of thinking about the systemic manifestations of dyadic phenomena and a new way of measuring the cohesion of the international system in terms of its Kantian fractionalization. These contributions further our understanding of the international system, providing new tools to consider the international system both theoretically and empirically. As a novel use of community detection in networks, Kantian fractionalization extends the multiplex network application of multislice modularity. Meanwhile, the resulting model for new violent interstate conflict, with Granger causality demonstrated forward up to seven years, may serve as an early-warning  signal of international instability. We hope that subsequent improvements to models utilizing Kantian fractionalization will further improve our understanding of both the dyad-level and system-level drivers of inter-state conflict.

\bibliographystyle{Science}
\bibliography{modularity}

\begin{thebibliography}{10}

\bibitem{Kossinets:2006}
G.~Kossinets, D.~J. Watts, {\it Science\/} {\bf 311}, 88–90 (2006).

\bibitem{Onnela:2007}
J.~Onnela, {\it et~al.\/}, {\it Proceedings of the National Academy of
  Sciences\/} {\bf 104}, 7332–7336 (2007).

\bibitem{Lazer:2009}
D.~Lazer, {\it et~al.\/}, {\it Science\/} {\bf 323}, 721–723 (2009).

\bibitem{Centola:2010}
D.~Centola, {\it Science\/} {\bf 329}, 1194–1197 (2010).

\bibitem{Bassett:2011}
D.~S. Bassett, {\it et~al.\/}, {\it Proceedings of the National Academy of
  Sciences\/} {\bf 108}, 7641 –7646 (2011).

\bibitem{Porter:2005}
M.~A. Porter, P.~J. Mucha, M.~E.~J. Newman, C.~M. Warmbrand, {\it Proceedings
  of the National Academy of Sciences\/} {\bf 102}, 7057 (2005).

\bibitem{Fowler:2010}
J.~H. Fowler, N.~A. Christakis, {\it Proceedings of the National Academy of
  Sciences\/} {\bf 107}, 5334 (2010).

\bibitem{MuchaPorter:2010}
P.~J. Mucha, M.~A. Porter, {\it Chaos\/} {\bf 20}, 041108–041108–1 (2010).

\bibitem{Maoz:1993}
Z.~Maoz, B.~Russett, {\it American Political Science Review\/} {\bf 87}, 624
  (1993).

\bibitem{Maoz:2006}
Z.~Maoz, {\it Journal of Peace Research\/} {\bf 43}, 391 (2006).

\bibitem{Ward:2007}
M.~D. Ward, R.~M. Siverson, X.~Cao, {\it American Journal of Political
  Science\/} {\bf 51}, 583 (2007).

\bibitem{Mucha:2010}
P.~J. Mucha, T.~Richardson, K.~Macon, M.~A. Porter, J.-P. Onnela, {\it
  Science\/} {\bf 328}, 876 (2010).

\bibitem{Kant:1795}
I.~Kant, {\it Perpetual Peace and Other Essays\/}, T.~Humphrey, ed. (Hackett,
  Indianapolis, IN, 1795), pp. 107--144.

\bibitem{Barbieri:1996}
K.~Barbieri, {\it Journal of Peace Research\/} {\bf 33}, 29 (1996).

\bibitem{Russett:1990}
B.~M. Russett, {\it Alternative Security: Living Without Nuclear Deterrence\/},
  B.~H. Weston, ed. (Westview Press, Boulder, CO, 1990), pp. 107--136.

\bibitem{Bennett:2003}
D.~S. Bennett, A.~C. Stam, {\it The Behavioral Origins of War\/} (University of
  Michigan Press, Ann Arbor, MI, 2003).

\bibitem{Russett:2001}
B.~M. Russett, J.~R. O'Neal, {\it {Triangulating Peace: Democracy,
  Interdependence and International Organizations}\/} (W. W. Norton \& Company,
  New York, NY, 2001).

\bibitem{Maoz:2011}
Z.~Maoz, {\it Networks of Nations: The Evolution, Structure, and Impact of
  International Networks: 1816-2001\/} (Cambridge University Press, New York,
  NY, 2011).

\bibitem{OnealRussett:1999}
J.~R. Oneal, B.~Russett, {\it World Politics\/} {\bf 52}, 1 (1999).

\bibitem{GleditschHegre:1997}
N.~P. Gleditsch, H.~Hegre, {\it The Journal of Conflict Resolution\/} {\bf 41},
  283 (1997).

\bibitem{CrescenziEnterline:1999}
M.~J.~C. Crescenzi, A.~J. Enterline, {\it Journal of Peace Research\/} {\bf
  36}, 75 (1999).

\bibitem{Porter:2009}
M.~A. Porter, J.~P. Onnela, P.~J. Mucha, {\it Notices of the American
  Mathematical Society\/} {\bf 56}, 1082–1097 \& 1164–1166 (2009).

\bibitem{Fortunato:2010}
S.~Fortunato, {\it Physics Reports\/} {\bf 486}, 75–174 (2010).

\bibitem{Newman:2004}
M.~E.~J. Newman, M.~Girvan, {\it Physical Review E\/} {\bf 69}, 026113 (2004).

\bibitem{GenLouvain}
I.~S. Jutla, L.~G.~S. Jeub, P.~J. Mucha, A generalized {L}ouvain method for
  community detection implemented in {MATLAB} (2011--2012).

\bibitem{Wymbs:2012}
N.~F. Wymbs, D.~S. Bassett, P.~J. Mucha, M.~A. Porter, S.~T. Grafton, {\it
  Neuron\/} {\bf 74}, 936–946 (2012).

\bibitem{Fortunato:2007}
S.~Fortunato, M.~Barth\'elemy, {\it Proceedings of the National Academy of
  Sciences\/} {\bf 104}, 36–41 (2007).

\bibitem{Good:2010}
B.~H. Good, Y.-A. de~Montjoye, A.~Clauset, {\it Physical Review E\/} {\bf 81},
  046106 (2010).

\bibitem{Reichardt:2006}
J.~Reichardt, S.~Bornholdt, {\it Physical Review E (Statistical, Nonlinear, and
  Soft Matter Physics)\/} {\bf 74}, 016110–14 (2006).

\bibitem{Marshall:2002}
M.~G. Marshall, K.~Jaggers, {\it Polity IV Project: Political Regime
  Characteristics and Transitions, 1800-2002\/}, Center for International
  Development and Conflict Management, University of Maryland, College Park,
  MD, version p4v2002e edn. (2002).

\bibitem{Leicht:2008}
E.~A. Leicht, M.~E.~J. Newman, {\it Physical Review Letters\/} {\bf 100},
  118703–4 (2008).

\bibitem{Jones:1996}
D.~M. Jones, S.~A. Bremer, J.~D. Singer, {\it Conflict Management and Peace
  Science\/} {\bf 15}, 163 (1996).

\bibitem{Cranmer:2012}
S.~J. Cranmer, B.~A. Desmarais, E.~J. Menninga, {\it Conflict Management and
  Peace Science\/} {\bf 23}, 279 (2012).

\bibitem{Maoz:2005}
Z.~Maoz, L.~G. Terris, R.~D. Kuperman, I.~Talmud, {\it New Directions for
  International Relations\/}, A.~Mintz, B.~Russett, eds. (Lexington, Lanham,
  MD, 2005), pp. 35--64.

\bibitem{Moul:1993}
W.~B. Moul, {\it Journal of Conflict Resolution\/} {\bf 37}, 735 (1993).

\bibitem{Ray:1973}
J.~L. Ray, J.~D. Singer, {\it Sociological Methods and Research\/} {\bf 1}, 403
  (1973).

\bibitem{Gelman:2007}
A.~Gelman, J.~Hill, {\it Data Analysis Using Regression and
  Multilevel/Hierarchical Models\/} (Cambridge University Press, New York, NY,
  2007).

\bibitem{Granger:1969}
C.~W. Granger, {\it Econometrica\/} {\bf 37}, 424 (1968).

\bibitem{Rubin:1974}
D.~Rubin, {\it Journal of Educational Psychology\/} {\bf 66}, 688 (1974).

\bibitem{Wayman:1985}
F.~W. Wayman, {\it Polarity and War: The Changing Structure of International
  Conflict\/}, A.~N. Sabrosky, ed. (Westview, Boulder, CO, 1985), pp. 93--111.

\bibitem{Wayman:1991}
F.~W. Wayman, T.~C. Morgan, {\it Measuring the Correlates of War\/}, P.~F.
  Diehl, J.~D. Singer, eds. (University of Michigan Press, Ann Arbor, MI,
  1991), pp. 139--158.

\bibitem{Singer:1972}
J.~D. Singer, S.~Bremer, J.~Stuckey, {\it Peace, War, and Numbers\/},
  B.~Russett, ed. (Sage, Beverly Hills, CA, 1972).

\bibitem{Singer:1987}
J.~D. Singer, {\it International Interactions\/} {\bf 14}, 115 (1987).

\bibitem{Blondel:2008}
V.~D. Blondel, J.-L. Guillaume, R.~Lambiotte, E.~Lefebvre, {\it Journal of
  Statistical Mechanics: Theory and Experiment\/} {\bf 2008}, P10008 (2008).

\bibitem{Macon:2012}
K.~T. Macon, P.~J. Mucha, M.~A. Porter, {\it Physica A\/} {\bf 391}, 343
  (2012).

\bibitem{Bassett:2013}
D.~S. Bassett, {\it et~al.\/}, {\it Chaos\/} {\bf 23}, 013142 (2013).

\end{thebibliography}

\baselineskip 24pt


\noindent \textbf{Acknowledgements:} The authors acknowledge a seed grant provided by the Howard W. Odum Institute for Social Science at the University of North Carolina at Chapel Hill.  PJM additionally acknowledges support from the NSF (DMS-0645369). SJC, EJM, \&  PJM conceived of the research and wrote the paper together. PJM performed the network community detection; SJC \& EJM performed the statistical analyses. The authors declare that they have no competing financial interests.


\newpage
\section*{Supporting Online Material for\\
\emph{Kantian Fractionalization Predicts the Conflict Propensity of the International System}}
   \setcounter{table}{0}\global\def\thetable{S\arabic{table}}%
   \setcounter{figure}{0}\global\def\thefigure{S\arabic{figure}}%

\begin{center}
Skyler J. Cranmer$^{1,\ast}$, Elizabeth J. Menninga,$^1$ and Peter J. Mucha$^2$\\ \bigskip
\normalsize{$^1$Department of Political Science, University of North Carolina, Chapel Hill, NC, USA}\\
\normalsize{$^2$Department of Mathematics, University of North Carolina, Chapel Hill, NC, USA}\\ \bigskip
\normalsize{$^\ast$To whom correspondence should be addressed; E-mail: skyler@unc.edu.}\\
\end{center}

\subsection*{Computation of Control Variables}

We included three control variables established in the literature to our models of conflict rate. Details on the computation of these variables are as follows.

\noindent \emph{Moul Polarity}\\
What we call Moul's polarity measure in the text is a modification of an earlier measure developed by Wayman \cite{Wayman:1985,Wayman:1991}. The Wayman measure is a proportion: the number of un-allied great powers plus the great power blocks formed by defensive alliances over the total number of great powers. So, in a given year $t$, Wayman's measure is computed as follows:

\[
\frac{\text{Un-allied Great Powers}_t + \text{Great Power Alliance Blocks}_t}{\text{Total Number of Great Powers}_t}.
\]

Moul gives the example that in 1950, the international system had five great powers divided into two groups, so Wayman's measure for 1950 is $2/5 = 0.40$ \cite[p.742]{Moul:1993}.  Moul then alters Wayman's measure by dividing each year's polarity by its minimum potential value. This makes values more comparable year-to-year. This measure sets $1$ as perfect bipolarization, and anything above $1$ as increasingly multipolar. In our data, we have one year of perfect (value $1$) bipolarity, one year at $1.5$, forty-one years at $2$, and ten years at $3$. 

\newpage

\noindent \emph{System Movement}\\
This measure captures changes in the international system's capabilities distribution. The specific measure we use is a five year moving average of the system movement variable established by Singer, Bremer, and Stuckey \cite{Singer:1972}. Movement is calculated as follows:
\[
MOVE = \frac{\sum_{i=1}^{N} | s_i^{t-1} - s_i^t | }{2 (1 - s_m^t)}
\]
where $N$ is the number of states in the system in a given year $t$,  $s_i$ is state $i$'s share of the international system's capabilities,  $s_m$ indicates the capability share of the state with the lowest share of capabilities, and the $t$ and $t-1$ superscripts indicate the time period in which the measure $s$ is taken. 

The share of a state's material capabilities is computed based on the Correlates of War National Material Capabilities dataset \cite{Singer:1972,Singer:1987}. Specifically, $s_i$ is state $i$'s share of the composite index of national capabilities (CINC), the most common operationalization of a state's power in International Relations. This index includes state $i$'s total population of country ratio (TPR), urban population of country ratio (UPR), iron and steel production of country ratio (ISPR), primary energy consumption ratio (ECR), military expenditure ratio (MER), and military personnel ratio (MPR). All ratios are taken as country over world. CINC is computed for state $i$ as: 
\[
CINC_i = \frac{TPR_i + UPR_i + ISPR_i + ECR_i + MER_i + MPR_i }{6}.
\]

\noindent \emph{Alliance Dependency}\\
We use, directly and without alteration, Maoz's measure of alliance interdependence\cite{Maoz:1993}. This measure is computed in several steps. 

First, the strength of commitment to the defense of a state is coded. Self-defense is seen as paramount, thus a state's relationship to itself is coded as 1. Defense pacts are coded as 0.75, non-aggression pacts are coded as 0.5, ententes are coded as 0.25, and a value of 0 is recorded if no alliance exists between two states. Based on these values, an $n \times n$ sociomatrix of alliance strength is created, where $n$ is the number of states in a given year. 

The values in the alliance strength matrix are then adjusted by the capabilities of the states involved in any given alliance. This is done because alliances are often seen as mechanisms for the aggregation of military capabilities. For example, suppose states $i$ and $j$ have a non-aggression pact (valued at 0.5), state $i$ has capabilities (measured as state $i$'s CINC score as described above) valued at 0.2, and state $j$ has capabilities valued at 0.007. State $i$ depends on state $j$ at level $0.007 \times 0.5 = 0.0035$, while state $j$ depends on state $i$ at level $0.2 \times 0.5 =  0.1$. This adjusted alliance dependency matrix, denoted as $A$ (but not to be confused with the $A_{ijl}$ Kantian adjacency elements used elsewhere in the present paper), then captures the \emph{direct} (first-order) dependences in the alliance network weighted by capabilities.

Because $A$ only captures direct dependencies, Maoz raises this matrix to the power of the number of degrees of dependence he wishes to capture. The alliance dependency of a given year's system of $N$ states is $A = \sum_{i=1}^{n-1} A^i + M$, where $n$ is the number of degrees of dependence and $M$ is the diagonal capability matrix such that $m_{ii}$ is the CINC score of state $i$ and non-diagonal entries are set to 0. 

Next, the row dependence of the matrix $A$ is computed as $a_{i.} = \sum_{j=1}^{n} a_{ij}$ and the total column dependence is computed as $a_{.i} = \sum_{i=1}^{n} a_{ij}$. The net dependence of a given state on others is then computed as $d_{i.} = (a_{i.} - a_{ij})/a_{i.}$ and the dependence of other states on any given state as $d_{.i} = (a_{.i} - a_{ij})/a_{.i}$. As Maoz describes: ``Finally, the overall strategic interdependence in the system is obtained by averaging the $d_{.i}$ row $(\bar{d}_{.i} = \frac{1}{n} \sum_{j=1}^{n} d_{,ij})$ and $d_{i.}$ $(\bar{d}_{i.} = \frac{1}{n}\sum_{j=1}^{n} d_{i.j})$ column of the matrix and averaging the two resulting averages''\cite[p.400]{Maoz:1993}.  Further details on the measurement of alliance dependency can be found in the original Maoz article.

\subsection*{Establishment of First Order Autocorrelation in Conflict Rate}

\begin{figure}
\centering
\begin{tabular}{cc}
\includegraphics[width=7cm]{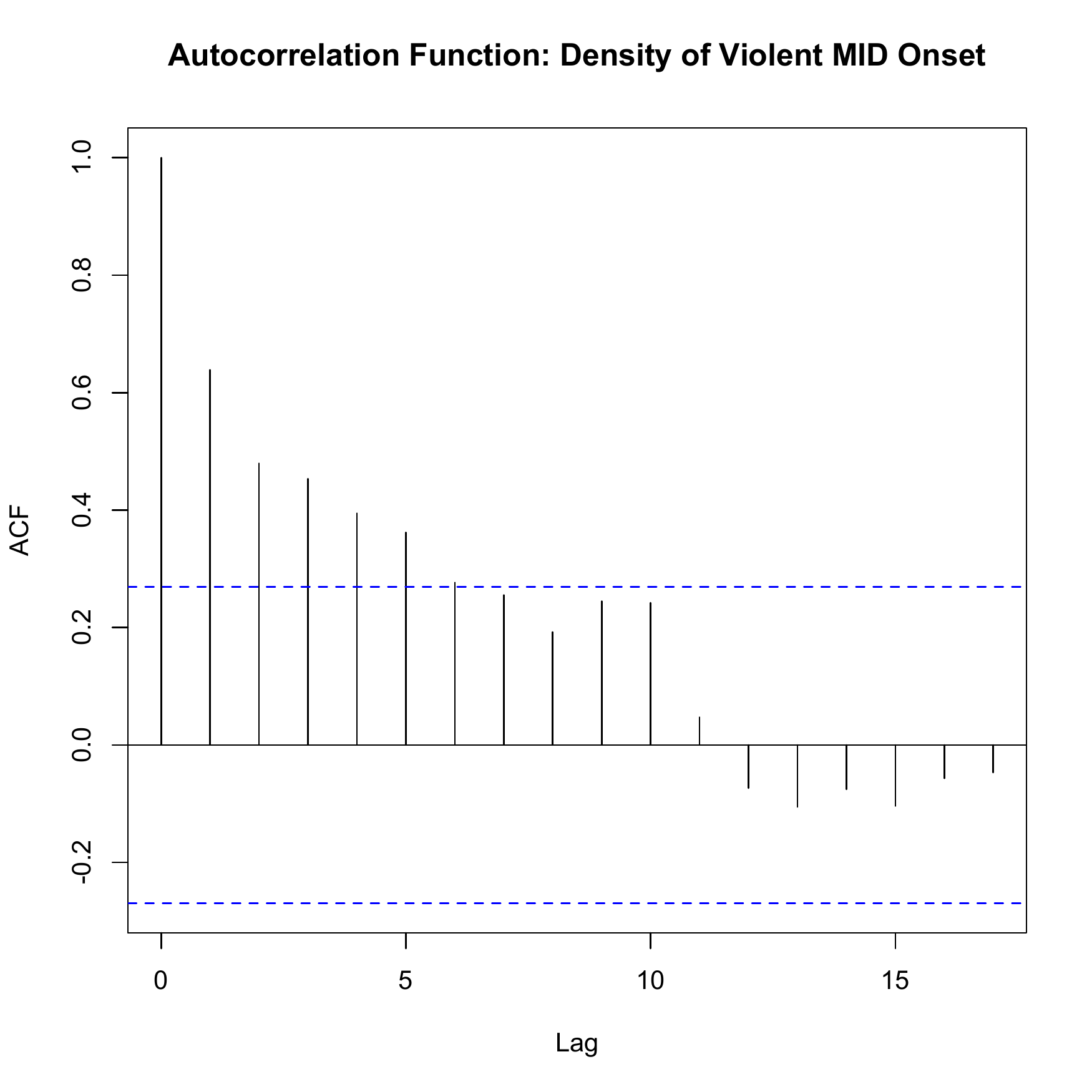}&
\includegraphics[width=7cm]{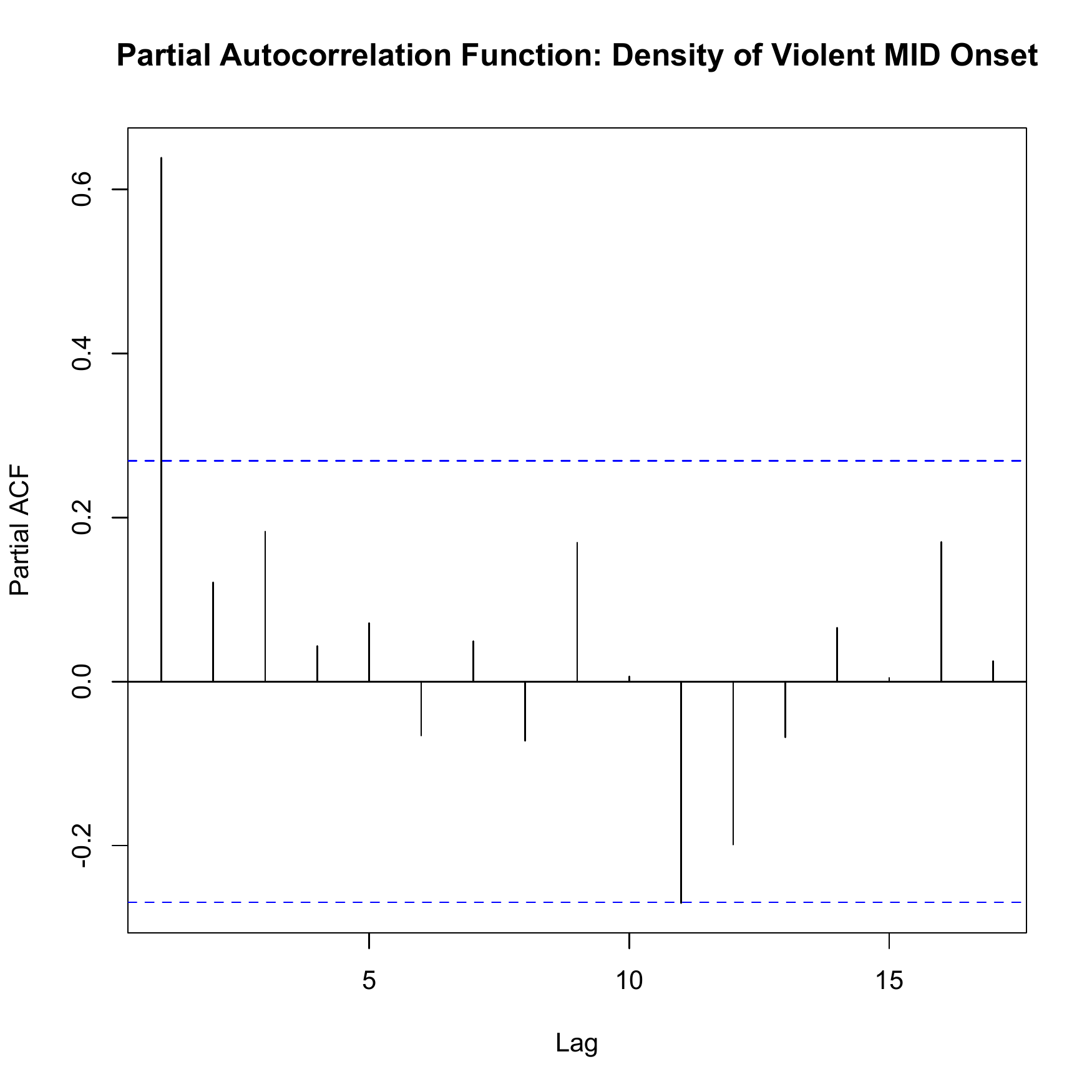}\\
\end{tabular}
\caption{Auto-Correlation Function and Partial Auto-Correlation Function Plots. 95\% confidence intervals are shown in blue. }
\label{ACF}
\end{figure}

The auto-correlation function (ACF) and partial auto-correlation function (PACF) plots in Fig. \ref{ACF} indicate first-order autocorrelation in the conflict rate variable, our outcome variable for most of the statistical analyses. The autocorrelation plot (left cell) shows the correlation of progressively lagged values of the variable with its contemporaneous values. First order autocorrelation typically produces the sort of steady decay out of significance that we observe in the plot. 

The PACF plot (right cell) is computed by fitting autoregressive models with progressively lagged values of the variable. First order autocorrelation typically produces the single spike followed by insignificant fits that we observe. Furthermore, the Breusch-Godfrey serial correlation Lagrange multiplier test, when applied to our linear specifications, confirms that autocorrelation is indeed first-order, but not higher (LM test = 52, $p$ = $5.109\cdot 10^{-12}$ for order 0, LM test = 0.055, $p$ = 0.815 for order 1).

Based on these plots, we concluded that the conflict rate variable is first-order autoregressive and we adjusted our statistical models accordingly by including one period (one year) lags of conflict rate on the right-hand-side of our models.

\subsection*{Rescaling Network Slices} The Kantian tripod as a multiplex network of common IGO membership, joint democracy, and interstate trade is represented (in each year) by valued weights in the adjacency elements $A_{ijl}$ describing the connections between countries $i$ and $j$ in layer $l$ (IGO, joint democracy, trade). Recognizing the differing units of weight in each layer---IGO links count the number of common memberships, joint democracy is a binary $\{0,1\}$ indicator between countries, and trade is defined by way of logarithms of real dollar values (details below)---a common scale was selected \emph{a priori} with the aim of putting these three kinds of connections on a comparable footing, so that the median present edge weight of each layer (considering the distribution of each across all time considered here) was set to be the same unit value. 

Under this selected rescaling, the joint democracy indicators remain as binary $\{0,1\}$ indicators. All common IGO membership weights have been rescaled by the obtained median of non-zero common membership counts across the 1948--2000 time period studied (median = 20 common memberships). The directed trade between countries $i$ and $j$ are similarly rescaled by the observed median weight of present edges, after a logarithmic transformation to minimize the dominance by the heavy tails in the trade distribution, as described in detail next. Because these normalizations to unit median present edge weights utilized a particular time period (1948--2000), small differences should be expected after a similar rescaling procedure applied to a different time window; nevertheless, we posit that the long time window yields a reasonable relative measurement of the roles of the three components of the Kantian tripod.

Defined in terms of real dollar values traded, the directed quantities of trade are heavy-tailed, with a handful of strong trade links far outweighing the other trade relationships in terms of dollars. This is in stark contrast to the joint democracy network layer used, which is a unit-weight clique between democratic states. Meanwhile, the weights given by numbers of common IGO memberships are more symmetrically distributed around their median and much closer to a Gaussian distribution than real dollars of trade. As comparisons, the ratio of maximum to median non-zero common IGO memberships in this data is $5.35$, while the ratio of the mean weight to the median is $\doteq 1.08$. In contrast, the ratio of maximum to median non-zero real dollar value of trade is $\doteq 1.1\cdot 10^5$, while the ratio of the mean dollar value to the median is $\doteq 70.6$.

In response to trade's heavy-tailed distribution in terms of real dollars, we defined the trade network in the Kantian tripod in terms of logarithms of real dollar values. This ensures that higher trade volumes do indeed receive heavier weights, while providing a more appropriate scale throughout. Further, we elect to linearly transform the logarithm of trade so that all non-zero trade values are positive and the rescaled log-trade weights have the same ratio between their minimum non-zero and median values (where the median is again taken over the whole distribution of non-zero values across the  time period studied) as the original dollar values of trade. That is, we identified the minimum, $v_\mathrm{min}$, and median, $v_\mathrm{median}$, of all non-zero real dollar trade values over all years studied (1948--2000). We then defined the logged-trade adjacency weight $A_{ij}$ between states $i$ and $j$ in terms of the observed non-zero real dollar value traded $v_{ij}>0$ as
\[
A_{ij} = r + (1-r)\frac{\log{(v_{ij}/v_\mathrm{min})}}{(-\log r)} 
\]
where $r=v_\mathrm{min}/v_\mathrm{median}$, with $A_{ij}=0$ if $v_{ij}=0$. Notably, because of the ratio of logarithms, the resulting edge weight is independent of the base of the logarithm. By construction, the rescaled edge weights have median $1$. The observed mean $\doteq 1.01$, and the maximum $\doteq 2.52$.

\subsection*{Robustness to coupling and resolution parameter choices} 

In the main text, we define Kantian fractionalization, $Q_K$, using default values $\gamma=\omega=1$ for the identity coupling strength $\omega$ and spatial resolution parameter $\gamma$. At these parameters, each realization of the selected modularity-optimizing heuristic (a generalized\cite{GenLouvain} Louvain\cite{Blondel:2008} algorithm) identifies between 2 and 6 communities in the Kantian tripod representation of the international system. While the resulting assignments might be interesting in their own right, we restrict our attention here to the measure of fractionalization provided by $Q_K$ itself. Nevertheless, when faced with such resolution parameter choices, we are cautious about the possible impact of making different choices. Strategies for letting the data guide resolution parameter selection include seeking roughly constant numbers of communities along some plateau in the parameter plane \cite{Macon:2012} or post-optimization null model testing for the statistical significance of the identified modularity at each parameter selection point \cite{Bassett:2013}. Given both the computational intensity and the need for a suitable random graph model for comparison in the latter approach, we concentrate our attention on the simplicity of the former approach, considering  $\gamma$ and $\omega$ separately. One could additionally consider different values of $\omega$ for individual states or different values of $\gamma$ for each multiplex network layer. However, in the absence of additional information that we might use to guide such choices, we restrict our attention here to the simplest uniform choices for these parameters.

In running the community detection for different $\omega\in [0,4]$ (with $\gamma=1$), we did not identify any dominant effect other than the expected significant difference between non-zero $\omega$ and the degenerate $\omega=0$ results (where the multislice representation of the multiplex Kantian tripod is not connected). In the absence of a data-led value for $\omega$, we compared our model specification for Kantian fractionalization at default resolutions in the main text with an alternative model based in terms of a $Q$ value obtained as an average over multislice modularities calculated in $\omega\in [0,4]$ (with $\gamma=1$), in steps of 0.1. The results are presented in Tables \ref{reg_table_means} and \ref{granger_means}.

\begin{table}[t]
\begin{center}
\begin{footnotesize}
\begin{tabular}{rrlrlrlrl}
 & \multicolumn{2}{c}{Linear Model 1} & \multicolumn{2}{c}{Linear Model 2} & \multicolumn{2}{c}{Count Model 1} & \multicolumn{2}{c}{Count Model 2}\\ 
  \hline
Kantian Fractionalization (lag) & \textbf{0.042} & \textbf{(0.012)} & \textbf{0.054} & \textbf{(0.019)} & \textbf{24.804} & \textbf{(3.447)} & \textbf{25.210} & \textbf{(5.790)} \\ 
Moul Polarity & -- & -- &  -0.000 & (0.001) & -- & -- & -0.166 & (0.203) \\ 
  Alliance Dependency (lag) & -- & -- & 0.007 & (0.004) & -- & -- & 1.978 & (1.224) \\ 
  System Movement (5 year) & -- & -- & 0.003 & (0.018) & -- & -- & -0.261 & (6.812) \\ 
  Conflict Density/Count (lag) &  \textbf{0.342} & \textbf{(0.132)} & 0.222 & (0.147) & \textbf{0.018} & \textbf{(0.005)} & \textbf{0.013} & \textbf{(0.006)} \\ 
   (Intercept)  & \textbf{-0.002} & \textbf{(0.001)} & -0.006 & (0.003) & \textbf{-8.549} & \textbf{(0.366)} & \textbf{-9.423} & \textbf{(1.053)}  \\ 
   \hline
   Adjusted $R^2$ / AIC & 0.52 & & 0.53 & & 365.56 & & 364.14 & \\
\end{tabular}
\caption{Kantian fractionalization computed by averaging over $\omega\in [0,4]$ (with $\gamma=1$). Linear models of violent conflict onset density and count models of violent conflict onset in the interstate system. Count models correct for over-dispersion and include the log of the number of dyads in the system year as an offset. Coefficients and standard errors displayed in bold are statistically significant at or below the  $p=0.05$ level.}
\label{reg_table_means}
\end{footnotesize}
\end{center}
\end{table}

\begin{table}
\centering
\begin{footnotesize}
\begin{tabular}{c|cc|cc}
& \multicolumn{2}{c}{Conflict Dens. $\rightarrow$ Kantian Frac.}\vline
& \multicolumn{2}{c}{Kantian Frac. $\rightarrow$ Conflict Dens.}\\
Lags & $F$-Statistic & $p$-Value & $F$-Statistic & $p$-Value\\
\hline
1 & 0.037 & 0.848 & \textbf{12.136} & \textbf{0.001} \\ 
  2 & 0.088 & 0.916 & \textbf{4.213} & \textbf{0.021} \\ 
  3 & 0.219 & 0.882 & \textbf{3.710} & \textbf{0.018} \\ 
  4 & 0.856 & 0.498 & \textbf{3.811} & \textbf{0.010} \\ 
  5 & 0.711 & 0.619 & \textbf{2.829} & \textbf{0.029} \\ 
  6 & 0.545 & 0.770 & 1.906 & 0.108 \\ 
  7 & 0.751 & 0.631 & \textbf{2.618} & \textbf{0.030} \\ 
  8 & 0.735 & 0.660 & 1.668 & 0.151 \\ 
  9 & 0.597 & 0.787 & 1.037 & 0.439 \\ 
  10 & 0.732 & 0.687 & 0.977 & 0.490 \\ 
  \hline
  \end{tabular}
\caption{Kantian fractionalization computed by averaging over $\omega\in [0,4]$ (with $\gamma=1$). Granger causal analysis of violent conflict onset density and Kantian fractionalization. Those $F$-statistics and $p$-values shown in bold are statistically significant at or below the  $p=0.05$ level.}
\label{granger_means}
\end{footnotesize}
\end{table}

Varying $\gamma$ has a strongly pronounced effect on the communities obtained, in that at $\gamma=0$ there is no partitioning into communities, with the numbers of communities increasing with $\gamma$ until by $\gamma=4$ almost every nation has been separately placed in a community by itself (with all three multislice nodes of that country in the multislice representation assigned to a group together). For $\omega=1$, we do not identify any plateau in the number of communities as $\gamma$ varies, so that this simple test does not identify a particular $\gamma$ resolution of interest for this system. As a  means of identifying some special value of $\gamma$, we consider the number of communities in a given year that have more than three multislice nodes assigned (that is, more than one country). While results vary from year-to-year, we identify a broad peak in the number of such communities near $\gamma=2$. Motivated by this weak indicator of a scale of interest, we additionally consider an alternative model in terms of $Q$ at $\gamma=2$ and $\omega=1$. The results are presented in Tables \ref{mult_fixed} and \ref{granger_fixed}.

Importantly, neither model based on alternative resolution parameter choices alters the qualitative result that the multiplex modularity of the Kantian tripod captures the fractionalization of the international system in a way that well models the onset of new conflicts.  Nevertheless, it certainly remains possible that statistical significance testing may yet uncover distinguished $(\gamma,\omega)$ parameter value regions that best uncover communities of nations (and of their Kantian tripod behaviors). Any future work investigating the specific assignments of such community detection in the Kantian tripod would do well to further explore the resolution parameter space for such regions. For the purposes of the present work, however, we have statistically established $Q_K$ as a measure of fractionalization (at the default $\gamma=\omega=1$) providing a useful quantity for modeling the rate of violent inter-state conflict. 

\begin{table}
\centering
\begin{footnotesize}
\begin{tabular}{rrlrlrlrl}
& \multicolumn{2}{c}{Linear Model 1} & \multicolumn{2}{c}{Linear Model 2} & \multicolumn{2}{c}{Count Model 1} & \multicolumn{2}{c}{Count Model 2}\\
\hline 
Kantian Fractionalization (lag) & \textbf{0.027} & \textbf{(0.008)} & \textbf{0.037} & \textbf{(0.013)} & \textbf{16.288} & \textbf{(2.315)} & \textbf{16.590} & \textbf{(3.835)} \\
Moul Polarity & -- & -- & -0.001 & (0.001) & -- & -- & -0.201 & (0.197) \\
Alliance Dependency (lag) & -- & -- & 0.008 & (0.004) & -- & -- & 2.260 & (1.230) \\
System Movement (5 year) & -- & -- & 0.003 & (0.018) & -- & -- & -0.198 & (6.757) \\
Conflict Density/Count (lag) & \textbf{0.355} & \textbf{(0.132)} & 0.213 & (0.148) & \textbf{0.019} & \textbf{(0.006)} & \textbf{0.013} & \textbf{(0.006)} \\
(Intercept) & -0.001 & (0.001) & -0.006 & (0.003) & \textbf{-8.096} & \textbf{(0.314)} & \textbf{-9.061} & \textbf{(0.995)} \\ \hline
Adjusted $R^2$ / AIC & 0.52 & & 0.53 & & 368.42 & & 364.63 & \\
\end{tabular}
\caption{Kantian fractionalization computed with $\gamma=2$ and $\omega=1$. Linear models of violent conflict onset density and count models of violent conflict onset in the interstate system. Count models correct for over-dispersion and include the log of the number of dyads in the system year as an offset. Coefficients and standard errors displayed in bold are statistically significant at or below the $p=0.05$ level.}
\label{mult_fixed}
\end{footnotesize}
\end{table}

\begin{table}
\centering
\begin{footnotesize}
\begin{tabular}{c|cc|cc}
& \multicolumn{2}{c}{Conflict Dens. $\rightarrow$ Kantian Frac.}\vline
& \multicolumn{2}{c}{Kantian Frac. $\rightarrow$ Conflict Dens.}\\
Lags & $F$-Statistic & $p$-Value & $F$-Statistic & $p$-Value\\
\hline
1 & 0.007 & 0.934 & \textbf{11.441} & \textbf{0.001} \\
2 & 0.052 & 0.949 & \textbf{3.922} & \textbf{0.027} \\
3 & 0.254 & 0.858 & \textbf{3.755} & \textbf{0.018} \\
4 & 0.827 & 0.516 & \textbf{4.657} & \textbf{0.004} \\
5 & 0.643 & 0.669 & \textbf{3.711} & \textbf{0.008} \\
6 & 0.387 & 0.882 & \textbf{2.692} & \textbf{0.030} \\
7 & 0.807 & 0.588 & \textbf{3.521} & \textbf{0.007} \\
8 & 0.687 & 0.700 & \textbf{2.385} & \textbf{0.042} \\
9 & 0.646 & 0.748 & 1.607 & 0.167 \\
10 & 0.687 & 0.725 & 1.357 & 0.263 \\ \hline
\end{tabular}
\caption{Kantian fractionalization computed with $\gamma=2$ and $\omega=1$. Granger causal analysis of violent conflict onset density and Kantian fractionalization. Those $F$-statistics and $p$-values shown in bold are statistically significant at or below the  $p=0.05$ level.}
\label{granger_fixed}
\end{footnotesize}
\end{table}

\subsection*{Identified Communities}
\begin{figure}
\begin{center}
\vspace{-1 in}
\includegraphics[width=\textwidth]{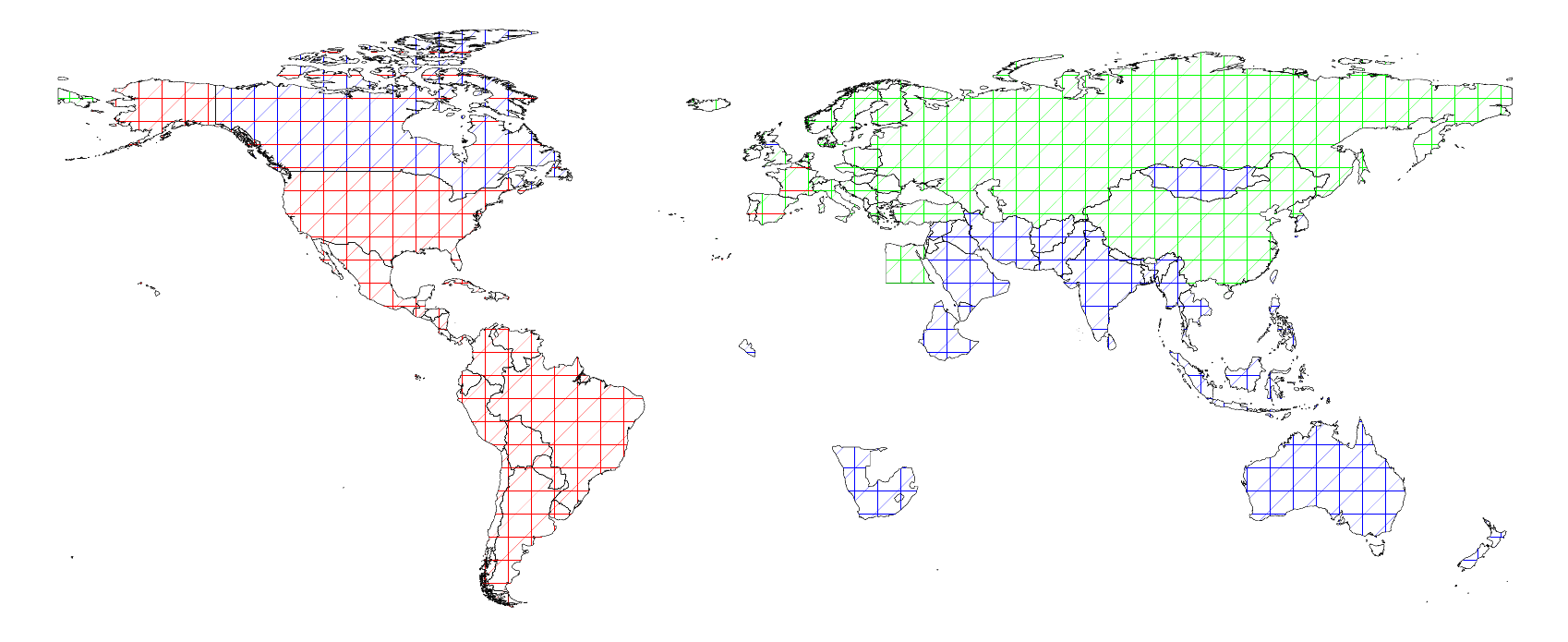}\\
\includegraphics[width=\textwidth]{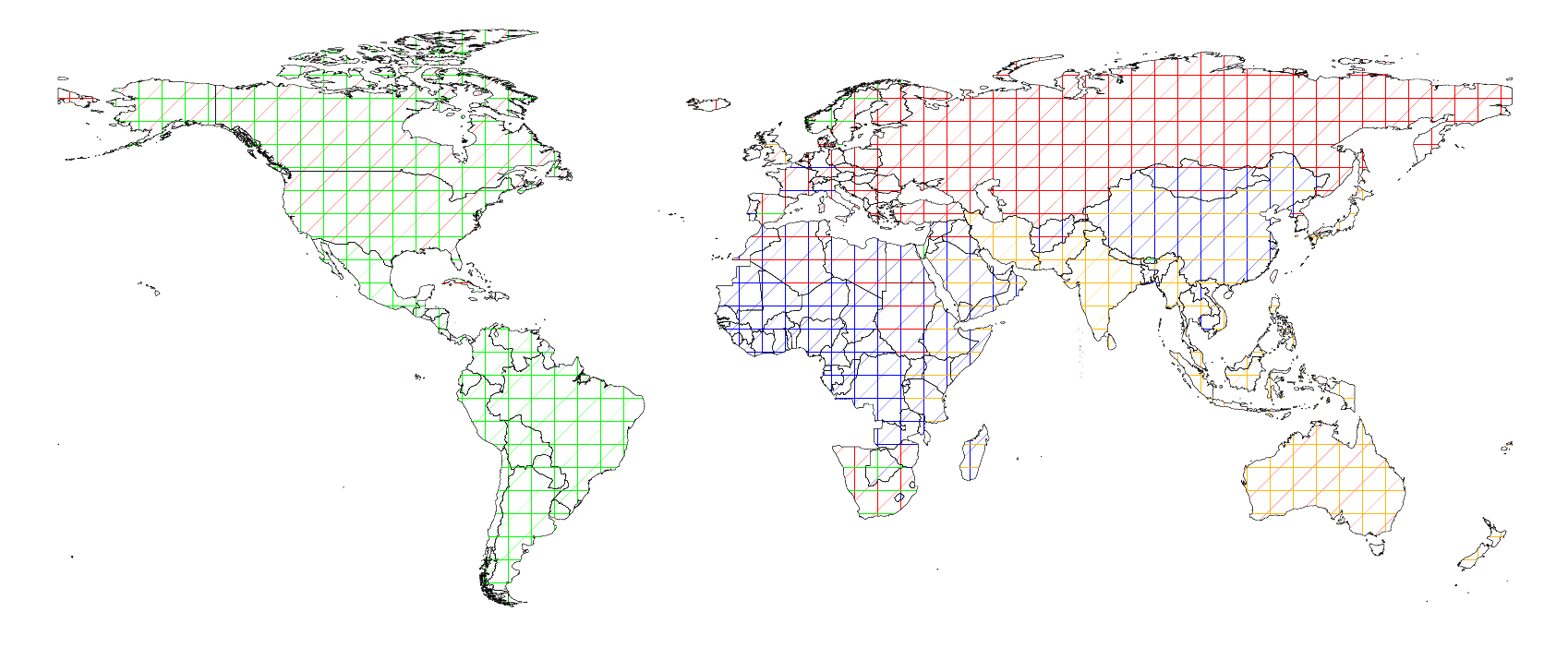}\\
\includegraphics[width=\textwidth]{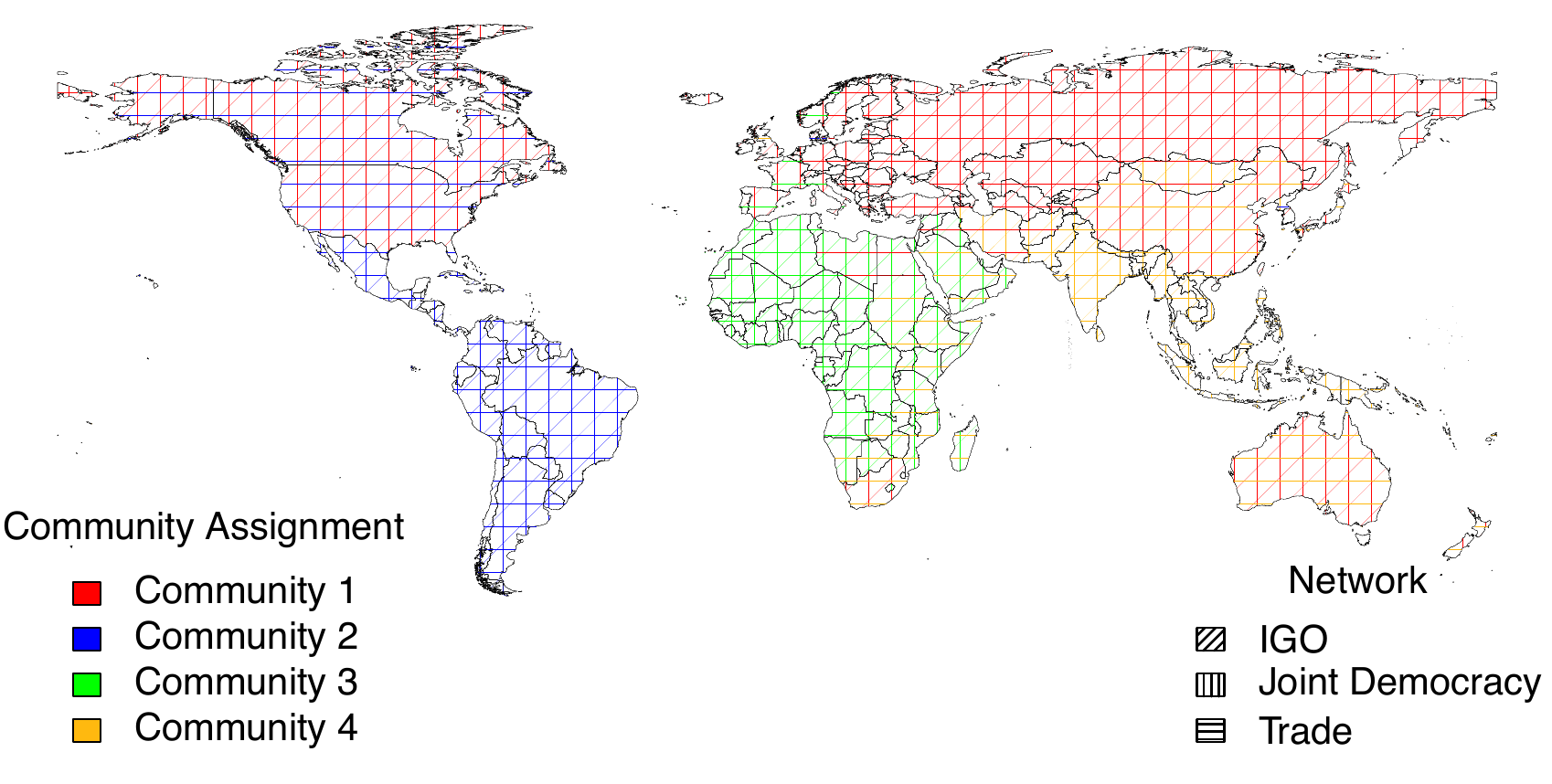}
\caption{Maps of the international system in 1950, 1975, and 2000 (top to bottom), with community assignments generated by a single realization of the computational heuristic \cite{GenLouvain} at default resolution parameters ($\gamma=\omega=1$). }
\label{maps}
\end{center}
\end{figure}

While we are not focused on any country's specific community assignments, we did visualize community assignments generated by our computation of Kantian fractionalization.  We present three of these visualizations below. Figure \ref{maps} illustrates the community assignments in the international system in 1950, 1975, and 2000, as generated by a single realization of the selected computational heuristic. (Interpretations of such assignments should be considered cautiously, however, since other realizations and heuristics can provide different assignments.) These maps outline all independent countries in the international system in the given year and then illustrate which communities each country belongs to. As each country is represented by 3 vertices (one for each layer of the Kantian tripod) each country could be in only 1 community (all 3 vertices placed together in the same community) or as many as 3 communities (each vertex placed differently). The years chosen represent the beginning, middle, and end of our time series.  Overall, these maps indicate that the community assignments used in our analysis  reflect known patterns of connectivity in the international system. This provides confidence that our community detection procedure did in fact identify meaningful communities based upon the Kantian tripod.

In 1950, this partition divides the international system into three communities. These communities predominantly fall into three geographic areas: the Americas, Europe and Asia, and  countries bordering the Indian Ocean. Most countries had all three of its vertices assigned to the same community. (Canada is a notably observable exception here, with strong IGO ties to the Indian Ocean rim.) In 1975 the picture is  more complicated. The observed communities still  break the international system into the Americas, Europe, and countries bordering the Indian Ocean, while identifying a fourth community that consists predominantly of countries  in Africa, but the breaks from these geographic divisions are more numerous. Community 2 includes China with Africa. Some countries in Europe are no longer assigned to only one community, and the United States shares a community with the Soviet Union in the  international organizations slice. These shifts reflect important changes to the international system as postwar organizations became important connections between the East and the West.  Europe maintained some connections with the Soviet Union, but many countries, as would be expected, shifted trade ties to the Americas or their newly independent colonies in Africa. In 2000, many countries continue to have divided community associations.  Notably the United States is still connected to the East through IGO membership, but maintains its strong assignment with the rest of the Americas through trade. China has moved from the IGO community dominated by African countries to join the US, Europe, and Russia. This reflects the strengthening of regional institutions in post-independence Africa as well as mainland China's increased participation in organizations such as the International Monetary Fund.

These maps also highlight some well-known insights about changes in the international system during our time period of study. In 1950, many African countries have not yet obtained independence from their colonial rulers. By 1975, most of the continent has become independent, with a few more countries obtaining independence by 2000. Between 1975 and 2000 we also see an increase in the number of countries in Eastern Europe, reflecting the states that emerged from the dissolution of the Soviet Union.  This illustrates the growth in the number of countries in the international system during our time series, highlighting the importance of accounting for the number of dyads in our analyses.

\subsection*{Additional Robustness Checks}
First, as mentioned in the main text, we ran linear and count models of \emph{conflict density} and \emph{conflict count} using lagged values of the single-slice modularities of the trade and IGO networks. Table \ref{IGO_trade} displays the result of this analysis, including trade and IGO modularities separately. In Table \ref{IGO_trade_combined} we consider the modularity of the IGO and trade networks together, including the density of democracies in the international system as well. We find that including Kantian fractionalization improves the model fit.

\begin{table}
\begin{footnotesize}
\begin{tabular}{rrlrlrlrl}
     & \multicolumn{2}{c}{Linear Model 1} & \multicolumn{2}{c}{Linear Model 2} & \multicolumn{2}{c}{Linear Model 3} & \multicolumn{2}{c}{Linear Model 4}\\
  \hline
IGO Modularity (lag) 		& \textbf{0.071} & \textbf{(0.021)} 	&  &  				& \textbf{0.081} & \textbf{(0.031)} 	&  &  				 \\ 
Trade Modularity (lag) 	& &  				& \textbf{0.046} & \textbf{(0.018)} 	&  &  				& \textbf{0.050} & \textbf{(0.025)} 	 \\ 
Moul Polarity  			&  &  				&  &  				& -0.001 & (0.001) 	& -0.001 & (0.001) 	 \\ 
Alliance Dependency (lag) &  &  			&  &  				& 0.003 & (0.004) 	& 0.006 & (0.004) 	 \\  
System Movement (5 year) &  &  			&  &  				& -0.006 & (0.021) 	& 0.032 & (0.017) 	 \\ 
Conflict Density (lag) & \textbf{0.368} & \textbf{(0.131)} 	& \textbf{0.500} & \textbf{(0.119)} 	& 0.265 & (0.144) 	& \textbf{0.376} & \textbf{(0.132)} 	 \\
(Intercept) 			& \textbf{-0.003} & \textbf{(0.001)} 	& -0.003 & (0.002) 	& -0.004 & (0.003) 	& -0.006 & (0.004) 	 \\ 
\hline
\vspace{0.5cm} Adjusted $R^2$ & 0.515 & & 0.475 & & 0.516 & & 0.489 &\\  
   
   & \multicolumn{2}{c}{Count Model 1} & \multicolumn{2}{c}{Count Model 2} & \multicolumn{2}{c}{Count Model 3} & \multicolumn{2}{c}{Count Model 4}\\
   \hline
IGO Modularity (lag) 			& \textbf{41.749} & \textbf{(6.567)} 	&  &  				& \textbf{36.285} & \textbf{(9.879)} 	&  &  \\ 
Trade Modularity (lag) 		&  &  				& \textbf{30.378} & \textbf{(6.074)} 	&  &  				& \textbf{26.013} & \textbf{(8.075)} \\ 
Moul Polarity  				&  &  				&  &  				& \textbf{-0.327} & \textbf{(0.187)} 	& \textbf{-0.419} & \textbf{(0.184)} \\ 
Alliance Dependency (lag) 	&  &  				&  &  				& 0.818 & (1.303) 	& \textbf{2.641} & \textbf{(1.320)} \\  
System Movement (5 year) 	&  &  				&  &  				& -2.833 & (7.574) 	& \textbf{15.331} & \textbf{(5.804)} \\ 
Conflict Count (lag) 			& \textbf{0.017} & \textbf{(0.006)} 	& \textbf{0.011} & \textbf{(0.006)} 	& \textbf{0.011} & \textbf{(0.006)} 	& 0.007 & (0.007) \\
(Intercept) 				& \textbf{-9.054} & \textbf{(0.482)} 	& \textbf{-8.982} & \textbf{(0.588)} 	& \textbf{-8.349} & \textbf{(0.987)} 	& \textbf{-9.634} & \textbf{(1.355)} \\ 
   \hline 
AIC  & 385.62 && 400.77 && 375.59 && 380.27 & \\
\end{tabular}
\caption{Using single-slide modularities of the Trade and IGO networks. Linear models of violent conflict onset density and count models of violent conflict onset in the interstate system. Count models correct for over-dispersion and include the log of the number of dyads in the system year as an offset. Coefficients and standard errors displayed in bold are statistically significant at or below the  $p=0.05$ level.}
\label{IGO_trade}
\end{footnotesize} 
\end{table}

\begin{table}
\begin{center}
\begin{footnotesize}
\begin{tabular}{l D{)}{)}{11)3} @{}D{)}{)}{11)3} @{}D{)}{)}{11)3} @{}D{)}{)}{11)3} @{}}

               & \multicolumn{1}{c}{Linear Model 1} & \multicolumn{1}{c}{Linear Model 2} & \multicolumn{1}{c}{Count Model 1} & \multicolumn{1}{c}{Count Model 2} \\
\midrule
Trade Modularity (lag)          & 0.039 \; (0.025)       & \textbf{0.052} \; \textbf{(0.025)} & \textbf{20.195} \; \textbf{(8.156)}   & \textbf{21.593} \; \textbf{(8.103)}   \\
IGO Modularity (lag)            & \textbf{0.058} \; \textbf{(0.023)} & \textbf{0.068} \; \textbf{(0.031)} & \textbf{29.530} \; \textbf{(7.172)}  & \textbf{26.741} \; \textbf{(10.276)}  \\
Democracy Density (lag) & 0.002 \; (0.005)       & 0.015 \; (0.011)      & 0.138 \; (1.730)         & 3.235 \; (3.896)         \\
Moul Polarity          &                        & -0.001 \; (0.001)     &                          & -0.364 \; (0.329)        \\
Alliance Dependency (lag)       &                        & 0.004 \; (0.004)      &                          & 1.205 \; (1.337)         \\
System Movement (5 year)       &                        & -0.013 \; (0.023)     &                          & -2.966 \; (8.376)        \\
Conflict Density/Count (lag)        & \textbf{0.304} \; \textbf{(0.133)}  & 0.158 \; (0.150)      &     \textbf{0.016} \; \textbf{(0.005)}                     &      \textbf{0.011} \; \textbf{(0.006)}                \\

(Intercept)      & \textbf{-0.006} \; \textbf{(0.002)} & \textbf{-0.008} \; \textbf{(0.004)} & \textbf{-10.087} \; \textbf{(0.810)} & \textbf{-10.152} \; \textbf{(1.407)} \\
  \hline
  Adjusted $R^2$ / AIC &  0.529  &  0.541 & 363.53  & 363.15  \\
\end{tabular}
\end{footnotesize}
\end{center}
\caption{Using single-slide modularities of the Trade and IGO networks. Linear models of violent conflict onset density and count models of violent conflict onset in the interstate system. Count models correct for over-dispersion and include the log of the number of dyads in the system year as an offset. Coefficients and standard errors displayed in bold are statistically significant at or below the  $p=0.05$ level. }
\label{IGO_trade_combined}
\end{table}

We conducted additional sets of robustness checks in order to verify that our variable of interest, Kantian fractionalization, is in fact driving the results of our statistical models. Table \ref{density_rob} displays the results of linear and count models that use the densities of the IGO, joint democracy, and trade networks (all lagged one year) as predictors of \emph{conflict density} and \emph{conflict count} (once again including a lagged dependent variable). This model assures us that the results found in Table \ref{reg_table} are not merely a function of the changes in density over time. 

\begin{table}
\begin{center}
\begin{footnotesize}
\begin{tabular}{l D{)}{)}{11)3} @{}D{)}{)}{11)3} @{}D{)}{)}{11)3} @{}D{)}{)}{11)3} @{}}

               & \multicolumn{1}{c}{Linear Model 1} & \multicolumn{1}{c}{Linear Model 2} & \multicolumn{1}{c}{Count Model 1} & \multicolumn{1}{c}{Count Model 2} \\
\midrule
Kantian Fractionalization       &                        & \textbf{0.074} \; \textbf{(0.027)} &                         & \textbf{35.284} \; \textbf{(8.251)} \\
IGO Density (lag)     & \textbf{-0.004} \; \textbf{(0.002)} & 0.003 \; (0.003)       & \textbf{-2.269} \; \textbf{(0.796)} & 1.504 \; (1.104)        \\
 Democracy Density (lag)   & 0.004 \; (0.006)       & -0.005 \; (0.006)      & 2.368 \; (2.313)        & -2.429 \; (2.230)       \\
Trade Density (lag)    & 0.003 \; (0.003)       & 0.000 \; (0.003)       & 1.308 \; (1.233)        & -0.775 \; (1.140)       \\
Conflict Density/Count (lag)       & \textbf{0.490} \; \textbf{(0.126)} & \textbf{0.299} \; \textbf{(0.137)}  &     \textbf{0.020} \; \textbf{(0.007)}                     &   \textbf{0.014} \; \textbf{(0.006)}   \\                    
(Intercept)    & 0.002 \; (0.002)       & \textbf{-0.007} \; \textbf{(0.004)}  & \textbf{-5.645} \; \textbf{(0.736)} & \textbf{-9.961} \; \textbf{(1.189)} \\
  \hline
  Adjusted $R^2$ / AIC & 0.449   & 0.515  & 408.86  & 366.94 \\
\end{tabular}
\end{footnotesize}
\end{center}
\caption{Linear and count models using densities rather than modularities as regressors. Count models correct for over-dispersion and include the log of the number of dyads in the system year as an offset. Coefficients and standard errors displayed in bold are statistically significant at or below the  $p=0.05$ level.}
\label{density_rob}
\end{table}

\begin{figure}[t]
\centering
\begin{tabular}{cc}\vspace{-0.5cm}
\includegraphics[width=8cm]{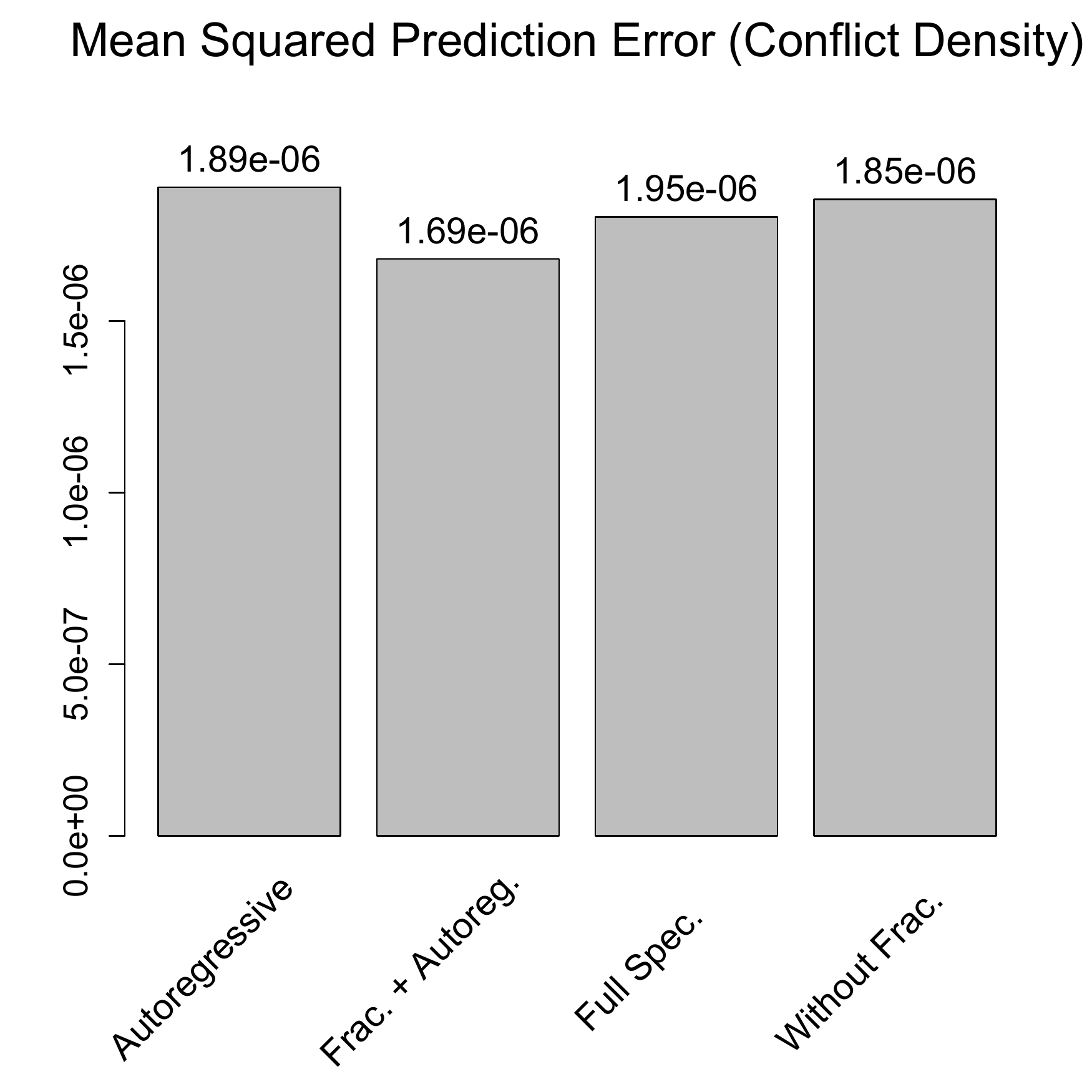}&
\includegraphics[width=8cm]{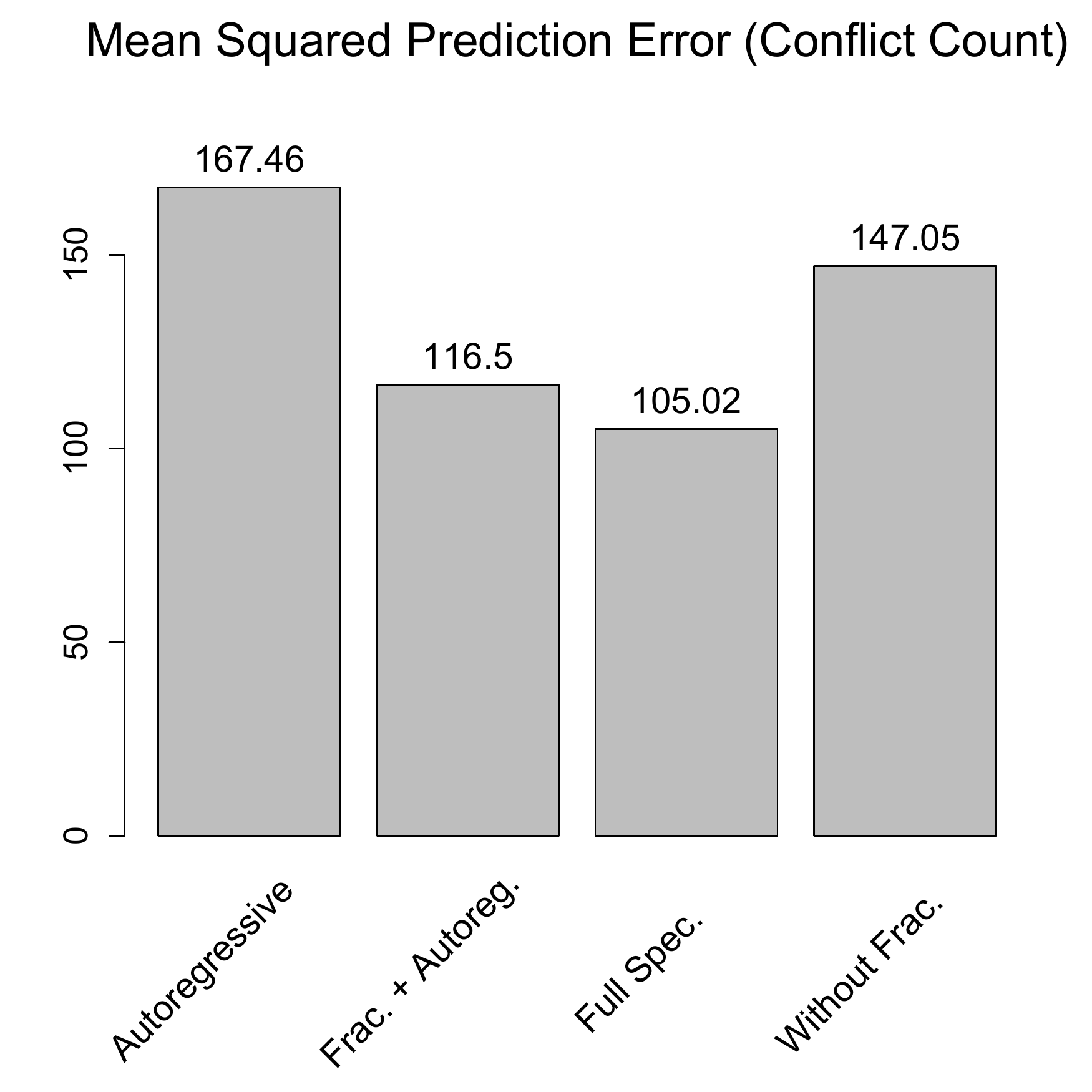}
\end{tabular}
\caption{Out-of-sample (one year ahead) predictive performance with Joint Democracy removed from the Kantian fractionalization computation. Both plots show the mean squared prediction error from a series of forecasts in which the values of both conflict density and conflict count, the left and right plots respectively, were forecast using only the data available up to, but not including, the year forecast.}
\label{prediction_nodem}
\end{figure}

Lastly, to verify the intuition from Fig. \ref{Qcontrib} in the main text, we recomputed Kantian modularity omitting joint democracy entirely. In other words, we computed the multislice modularity in the same way as before, but considering only the trade and IGO networks. We then re-ran the entire empirical analysis: regressions, Granger tests, and one-year-ahead out-of-sample prediction. The results from these analyses are presented in Tables \ref{nodem} and \ref{granger_nodem} and Fig.; \ref{prediction_nodem}. The results are very similar to those presented in the main text and none of the substantive conclusions we would draw from either set of results differ.

\begin{sidewaystable}
\begin{center}
\begin{footnotesize}
\begin{tabular}{l D{)}{)}{11)3} @{}D{)}{)}{11)3} @{}D{)}{)}{11)3} @{}D{)}{)}{11)3} @{}D{)}{)}{11)3} @{}D{)}{)}{11)3} @{}}
 
               & \multicolumn{1}{c}{Linear Model 1} & \multicolumn{1}{c}{Linear Model 2} & \multicolumn{1}{c}{Linear Model 3} & \multicolumn{1}{c}{Count Model 1} & \multicolumn{1}{c}{Count Model 2} & \multicolumn{1}{c}{Count Model 3} \\
\midrule
Kantian Fractionalization  (No-JD, lag)         & \textbf{0.064} \; \textbf{(0.017)}  & \textbf{0.082} \; \textbf{(0.026)} &                        & \textbf{35.644} \; \textbf{(4.874)} & \textbf{35.415} \; \textbf{(7.761)}  &                         \\
Moul Polarity        &                         & 0.000 \; (0.001)       & \textbf{-0.001} \; \textbf{(0.001)} &                         & -0.177 \; (0.195)        & \textbf{-0.726} \; \textbf{(0.165)} \\
Alliance Dependency (lag)     &                         & \textbf{0.007} \; \textbf{(0.004)}   & 0.003 \; (0.004)       &                         & \textbf{2.007} \; \textbf{(1.194)}     & 1.624 \; (1.455)        \\
System Movement (5 year)        &                         & 0.005 \; (0.017)       & \textbf{0.029} \; \textbf{(0.017)}   &                         & 0.996 \; (6.455)         & \textbf{15.177} \; \textbf{(6.147)}  \\
Conflict Density/Count (lag)       & \textbf{0.327} \; \textbf{(0.130)}   & 0.210 \; (0.142)       & \textbf{0.464} \; \textbf{(0.129)} &                        \textbf{0.017} \; \textbf{(0.005)} &              \textbf{0.013} \; \textbf{(0.006)}              &             0.006 \; (0.007)                   \\
(Intercept)    & \textbf{-0.003} \; \textbf{(0.001)} & \textbf{-0.008} \; \textbf{(0.004)} & 0.001 \; (0.002)       & \textbf{-9.285} \; \textbf{(0.455)} & \textbf{-10.124} \; \textbf{(1.131)} & \textbf{-5.891} \; \textbf{(0.804)} \\
\midrule
Adjusted $R^2$ / AIC      & 0.537                  & 0.545                  & 0.455                  &          362.48               &        361.27                 &   406.67                      \\
\end{tabular}
\end{footnotesize}
\end{center}
\caption{Kantian fractionalization computed without Joint Democracy. Count models correct for over-dispersion and include the log of the number of dyads in the system year as an offset. Coefficients and standard errors displayed in bold are statistically significant at or below the $p=0.05$ level. }
\label{nodem}
\end{sidewaystable}

\begin{table}
\centering
\begin{footnotesize}
\begin{tabular}{c|cc|cc}
& \multicolumn{2}{c}{Conflict Dens. $\rightarrow$ Kantian Frac.}\vline
& \multicolumn{2}{c}{Kantian Frac. $\rightarrow$ Conflict Dens.}\\
Lags & $F$-Statistic & $p$-Value & $F$-Statistic & $p$-Value\\
\hline
1 & 0.085 & 0.772 & \textbf{13.974} & \textbf{0.000} \\ 
  2 & 0.059 & 0.943 & \textbf{6.620} & \textbf{0.003} \\ 
  3 & 0.271 & 0.846 & \textbf{5.165} & \textbf{0.004} \\ 
  4 & 1.885 & 0.132 & \textbf{4.268} & \textbf{0.006} \\ 
  5 & 1.075 & 0.390 & \textbf{3.506} & \textbf{0.011} \\ 
  6 & 0.668 & 0.676 & \textbf{2.587} & \textbf{0.036} \\ 
  7 & 0.957 & 0.479 & \textbf{3.278} & \textbf{0.010} \\ 
  8 & 1.096 & 0.395 & 2.110 & 0.069 \\ 
  9 & 1.225 & 0.324 & 1.323 & 0.275 \\ 
  10 & 1.421 & 0.235 & 1.167 & 0.362 \\ 
  9 & 1.151 & 0.366 & 1.778 & 0.123 \\ 
  10 & 1.377 & 0.254 & 1.562 & 0.184 \\ \hline
\end{tabular}
\caption{Kantian fractionalization computed without Joint Democracy. Granger causal analysis of violent conflict onset density and Kantian fractionalization. Those $F$-statistics and $p$-values shown in bold are statistically significant at or below the $p=0.05$ level.}
\label{granger_nodem}
\end{footnotesize}
\end{table}

\end{document}